\documentclass[journal]{IEEEtran}
\ifCLASSINFOpdf
\else
\fi
\usepackage{longtable}

\usepackage{cite}

\usepackage{setspace}
\usepackage{amsmath}
\usepackage{amsthm, amssymb}
\usepackage{algorithm}
\usepackage[noend]{algpseudocode}
\makeatletter
\def\BState{\State\hskip-\ALG@thistlm}
\makeatother

\usepackage{float}
\usepackage{subfig}
\usepackage{graphicx}
\usepackage{mathrsfs}
\usepackage{romannum}

\newtheorem{theorem}{\textbf{Theorem}}
\newtheorem{fact}{\textbf{Fact}}

\newtheorem{lemma}{\textbf{Lemma}}

\newtheorem{corollary}{\textbf{Corollary}}

\theoremstyle{definition}
\newtheorem{definition}{\textbf{Definition}}

\newtheorem{example}{Example}

\DeclareMathOperator*{\argmax}{arg\,max}
\begin{document}
%
\title{Distributed Rumor Blocking with \\Multiple Positive Cascades}
%
%
%

\author{Guangmo (Amo)~Tong,~\IEEEmembership{Student Member,~IEEE,}
        Weili~Wu,~\IEEEmembership{Member,~IEEE,}
        and~Ding-Zhu~Du,
\thanks{G. Tong, W. Wu and D.-Z. Du are with the Department of Computer Science,  Erik
Jonsson School of Engineering and Computer Science, the University
of Texas at Dallas.}
}

%
%

\markboth{Journal of \LaTeX\ Class Files,~Vol.~13, No.~9, September~2014}%
{Shell \MakeLowercase{\textit{et al.}}: Bare Demo of IEEEtran.cls for Journals}
%



\maketitle

\begin{abstract}
Misinformation and rumor can spread rapidly and widely through online social networks and therefore rumor controlling has become a critical issue. It is often assumed that there is a single authority whose goal is to minimize the spread of rumor by generating a positive cascade. In this paper, we study a more realistic scenario when there are multiple positive cascades generated by different agents. For the multiple-cascade diffusion, we propose the P2P independent cascade (PIC) model for private social communications. The main part of this paper is an analysis of the rumor blocking effect (i.e. the number of the users activated by rumor) when the agents non-cooperatively generate the positive cascades. We show that the rumor blocking effect provided by the Nash equilibrium will not be arbitrarily worse even if the positive cascades are generated non-cooperatively. In addition, we give a discussion on how the cascade priority and activation order affect the rumor blocking problem. We experimentally examine the Nash equilibrium of the proposed games by simulations done on real social network structures. 
\end{abstract}


%
\IEEEpeerreviewmaketitle

\section{Introduction}
\IEEEPARstart{W}{ith} the recent advancements of information technologies, social networks have significantly changed the world by allowing efficient interchange of ideas and innovations. Especially in online social networks of which there is a drastic usage in the past decade, the hitting news may break out even before officially announced \cite{nguyen2012containment}. However, misinformation or rumor also spreads through the network \cite{gupta2013faking}, which may lead serious public panic or economic consequence. Therefore, rumor control has become one of the important issues in social networks research.  

The topics regarding rumor control are closely related to the study of influence diffusion in social networks. In a social network, it is assumed that information spreads in the fashion of influence cascades. Under the classic models, a cascade starts to spread from a set of seed users and then propagates from active users to inactive users. Rumor is taken as a certain cascade spreading along with other cascades, and the cascades holding opposite opinions may compete against each other. In particular, a user who has received the genuine news will not accept the rumor. Conversely, when rumor comes first, the undesirable effect can be caused immediately and therefore the true fact arriving later is futile. For example, when affected by the misinformation of swine flu on Twitter, people might have taken mistake vaccines before receiving the clarification from WHO. In order to prevent people from being misled by rumor, a natural method is to introduce a positive cascade that is able to reach users before the arrival of rumor. Once the rumor is detected, the network manager can generate a competing positive cascade by selecting appropriate seed users such that the number of rumor-activated users can be minimized. Motivated by this framework, several works (e.g., \cite{fan2013least, budak2011limiting, he2012influence, tong2017efficient}) have studied the rumor blocking problem under the competitive diffusion models.

The feasibility of the existing work is limited by the following aspects. On one hand, due to the great magnitude of a social network, the whole network cannot be efficiently controlled by a single manager. In a more realistic scenario, there are usually more than one positive cascades generated by different users or institutes which we call the \textit{agents}. Although all fight against rumor, when designing rumor containment strategies, such agents do not cooperate with each other. In this case, the rumor blocking task is distributed to the agents. Specifically, each agent makes their own choice according to the actions of other agents such that their own utility can be maximized, which forms a game between the agents.  Under this setting, the social objective is to minimize the number of rumor-activated users while the private utility of each agent varies under different games. In this paper, we study such a non-cooperative rumor blocking game and investigate the problem that \textit{how bad can the equilibrium of the game be in the worst-case compared with the optimal seeding strategy, with respect to the number of non-rumor-activated nodes}.    

By extending the classic independent cascade model, we herein develop the peer-to-peer independent cascade (PIC) model supporting the multiple-cascade diffusion. Unlike the existing models, the PIC model assumes that an active user can only activate one inactive neighbor at each time step. The PIC model represents the private social communication where the content is not open to all the users in the network. One example is the mobile social network where the communication is established by mobile phones in a person-to-person manner. Based on the PIC model, we formulate the rumor blocking game with one cascade of rumor and $k$ agents where each agent generates one positive cascade. In such a game, the social utility is the number of rumor-activated nodes, which is a function over the strategy space of the agents. In this paper, we first show that under the PIC model the social utility is a \textit{set function} of the union of the seed sets of the positive cascades, and furthermore, it is monotone increasing and submodular. For the private utility, we consider two games, the rumor-aware game and the rumor-oblivious game, depending on whether or not the agents are able to distinguish the rumor from the genuine news. For the proposed games, we provide an analysis of the equilibrium under the best-response assumption and the approximate-response assumption, respectively. Under the former, the agents are able to make optimal decisions and the equilibrium of the game provides a 2-approximation with respect to the social utility. In another issue, we consider that case that the agents cannot obtain the optimal strategy in polynomial time due to the NP-hard nature of the problem. As shown later, the private utility is submodular and it is well-known that the submodular maximization problem admits an efficient $(1-1/e)$-approximation. Assuming that the agents adopt such an approximation strategy, we prove that the equilibrium of the rumor blocking game provides a $\frac{2 e-1}{e-1}$-approximation. We simulate the rumor blocking game on graphs extracted from real-world social networks and record the number of nodes influenced by rumor. The experimental results have shown that the effect of non-cooperative rumor blocking game is comparable to that of the single-positive-cascade case when the seed nodes are selected by the state-of-the-art algorithms. 

In addition to the game-theoretic analysis, we further discuss the property of the completive diffusion models. When developing such kind of models, there are two critical settings. One is to determine which cascade should a user $u$ select when multiple cascades reach $u$ at the same time. Another one is the order of activation. That is, when a node becomes active, which of its neighbor will be firstly selected for activation. The activation order of the neighbors plays an important role in the diffusion of multiple cascades because the node is activated by the first cascade reaching it. As discussed later in Sec. \ref{sec:further}, such issues become tricky and complicated when there are more than two cascades. For example, under certain reasonable settings, when more positive cascades appear in the network, the rumor may paradoxically spread more widely\footnote{As discussed in prior works e.g. \cite{budak2011limiting} and \cite{chen2016centralized}, such issues become less important for the classic IC model because the influence spreads from one active node to all of their  neighbors simultaneously.}. In this paper, we will discuss such issues and provide several interesting observations on the property of competitive diffusion model.

\textbf{Contribution}. The contribution of this paper is summarized as follows.
\begin{itemize}
\item We propose a new competitive cascade model which represents the private peer-to-peer communication in social systems.
\item We formulate the rumor blocking game and provide the analysis of the equilibrium regarding the effect of rumor blocking. The main result is that the rumor blocking effect can be guaranteed with a provable ratio even if the agents work non-cooperatively.
\item We discuss the property of the competitive cascade model under different settings of cascade priority and activation order.
\end{itemize}

\textbf{Organization.} The rest of the paper is organized as follows. In Sec. \ref{sec:related} we survey the related work. In Sec. \ref{sec:pre}, we provide the preliminaries and formulate the PIC model. A game-theoretic analysis is given in Sec. \ref{sec: game}. In Sec. \ref{sec:further}, we discuss the property of the competitive cascade model under different settings. The experimental results are shown in Sec. \ref{sec:exp}. Sec. \ref{sec:con} concludes. Most of the proofs are given in Appendix \ref{sec: proof}.

\section{Related Work}
\label{sec:related}
Rumor control has drawn significant attention from both academia and industry. In what follows, we briefly introduce the prior works related to this topic.

Rumor detection aims to distinguish rumor from genuine news. Leskovec \textit{et al.} \cite{leskovec2009meme} develop a framework for tracking the spread of misinformation and observe a set of persistent temporal patterns in the news cycle. Ratkiewicz \textit{et al. }\cite{ratkiewicz2011detecting} build a machine learning framework to detect the early stages of viral spreading of political misinformation. In \cite{qazvinian2011rumor}, Qazvinian \textit{et al.} address this problem by exploring the effectiveness of three categories of features: content-based, network-based, and microblog-specific memes. Takahashi \textit{et al.} later \cite{takahashi2012rumor} study the characteristics of rumor and design a system to detect the rumor on Twitter. 

Rumor source detection is another important problem for rumor control. The prior works primarily focus on the classic susceptible-infected-recovered (SIR) model where the nodes can be infected by rumor and may recover later. Shah \textit{et al.} \cite{shah2011rumors} provide a systematic study and design a rumor source estimator based on the concept of rumor centrality. Z. Wang \textit{et al.} \cite{wang2014rumor} later study this problem with the consideration of multiple observations. 

The rumor blocking problem is mainly considered under the influence-propagation models. The study of influence diffusion can be tracked back to Domingos \textit{et al.} \cite{domingos2001mining}. Later in the seminal work of Kempe \textit{et al.} \cite{kempe2003maximizing}, two basic operational models, Independent Cascade model (IC) and Linear Threshold model (LT), are proposed. Based on those models, advanced models supporting multiple cascades are then developed and the competitive influence diffusion problem has been studied in such models. Bharathi \textit{et al.} \cite{bharathi2007competitive} show a $(1-1/e)$-approximation algorithm for the best response to an opponent's strategy. Borodin \textit{et al.} \cite{borodin2010threshold} study several competitive diffusion models by extending of the classic LT model and show that the original greedy approach proposed in \cite{kempe2003maximizing} may not be applicable to such settings. The rumor blocking problem is similar but not identical to the competitive influence maximization problem. The goal of the competitive influence maximization problem is to maximize the spread of a certain cascade while rumor blocking aims to minimize the spread of rumor (i.e. minimize the number of rumor-activated nodes). For the rumor blocking problem, Xinran \textit{et al.} \cite{he2012influence} show a $(1-1/e)$-approximation algorithm for the competitive Linear Threshold Model, and, Lidan \textit{et al.} \cite{fan2013least} study this problem under the OPOAO model and DOAM model. From another perspective, Nguyen \textit{et al.} \cite{nguyen2012containment} propose the $\beta_{T}^{I}$-Node Protector problem which limits the spread of misinformation by blocking the high influential nodes. Z. He \textit{et al.} \cite{he2017cost} study the rumor blocking problem in mobile social networks.
The above works all aim to design seeding algorithms, which is essentially different from the topic of this paper. 

We are not the first who study the influence diffusion via game theoretical approaches. Kostka \textit{et al.} \cite{kostka2008word} formulate the seeding process as a game and study the best-response strategy under a new model which is more restricted than the IC and LT model. Different from that paper, we do not design response strategies and instead our analysis focuses on the equilibrium of the game where there is one rumor cascade and multiple positive cascades. In another issue, C. Jiang \textit{et al.} in \cite{jiang2014evolutionary,jiang2014graphical} propose an evolutionary game theoretic framework to model the dynamic information diffusion process in social networks. 

\section{Model}
\label{sec:pre}
In this section, we introduce the system model and provide the preliminaries. The notations that are frequently used in this paper are listed in Table \ref{table:notations}.

\subsection{Influence Diffusion}
\label{subsec:model}

\subsubsection{Single Cascade}
A social network is given by a directed graph $G=(V,E)$ where $V$ and $E$ denote the users and social ties, respectively. Let $N_u=\{v|(u,v)\in E \}$ be the set of the out-neighbors of node $u$ and define $d_u=|N_u|$ as the number of the out-neighbors of $u$. We will use terms \textit{user} and \textit{node} interchangeably. We speak of each user as being \textit{active} and \textit{inactive}. To trigger the spread of influence, some users are firstly activated as seed users who will later attempt to activate their out-neighbors. Under the independent cascade model, associated with each edge $(u,v)$ there is a propagation probability $p_{(u,v)}^G \in [0,1]$ which is the probability that $u$ successfully activates $v$. For each pair of nodes $u$ and $v$, $u$ has only one chance to activate $v$. By the fashions of influence propagation, the independent cascade model can be classified into the following two categories. 

\textbf{Broadcast Independent Cascade (BIC) model.\footnote{This is the model that has been considered in most of the prior works \cite{kempe2003maximizing,zhu2015better,ok2016maximizing,lu2016towards}.}}
Under this model, when node $u$ becomes active at time $t-1$, it attempt to activates all of its out-neighbors simultaneously at time step $t$.

\begin{figure}[t]
\centering
\captionsetup{justification=centering}
\subfloat[BIC model.]{\label{fig:example_bic}\includegraphics[width=0.48\textwidth]{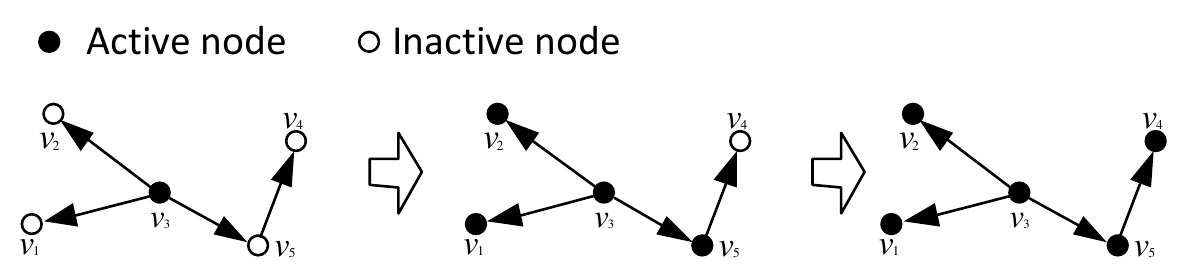}} \hspace{0mm}

\subfloat[PIC model.]{\label{fig:example_pic}\includegraphics[width=0.48\textwidth]{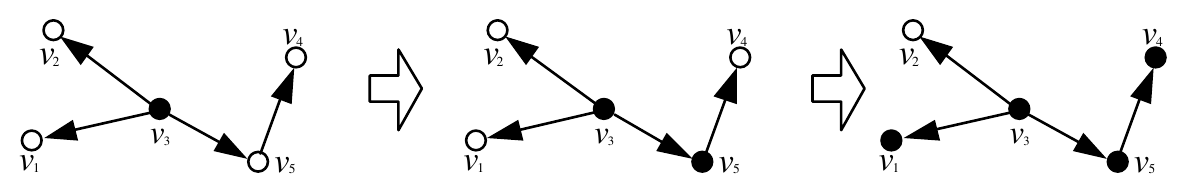}} \hspace{0mm} 
\caption{\small \textbf{An illustrative example of BIC and PIC models.}} 
\label{fig:example}
\end{figure}

\textbf{P2P Independent Cascade (PIC) model.} Under this model, an active node $u$ can only attempt to activate one of its out-neighbors at a time step.

\begin{example}
An illustrative example of the above two models is shown in Fig. \ref{fig:example}. Suppose the propagation probability of each edge is 1 and node $v_3$ is selected as the seed node. As shown in the figures, after the first step, all the neighbors of $v_3$ are activated under the BIC model, while only one neighbor of $v_3$ is activated under the PIC model.
\end{example}

The BIC model represents the open social communication namely Facebook or Twitter. For example, a public post on Facebook is simultaneously available to all the user's friends. The PIC model represents the private social communication such as personal online message or email, where a user has to take an action to pass the message to their friends. Note that, for the rumor blocking problem, there is a significant difference between these two models. One can see that the PIC model tends to slow the spread of influence, and when multiple cascades exist whether a node will be rumor-activated depends on the first cascade reaching it. In this paper, we focus on the PIC model which has not been studied in the literature.

\begin{table}[t]
\centering
{\begin{tabular}{ |p{2.1cm} || p{5.85cm} |}
\hline 
\textbf{Symbol}& \textbf{Definition} \\
\hline 
$G=(V,E)$& a PIC network.    \\
\hline
$N_u$ & the set of out-neighbors of node $u$.    \\
\hline
$p_{(u,v)}^G$& propagation probability of edge $(u,v)$.    \\
\hline
$C_r$& the cascade of rumor.    \\
\hline
$a_r$& the seed set of rumor.    \\
\hline
$b_u^t$& the neighbors of $u$ that can be activated by $u$ at time step $t$.    \\
\hline
$\mathrm{Pr}[g]$& the probability that the realization $g$ can be generated.    \\
\hline
$\mathcal{G}$& the set of all possible realizations.    \\
\hline
$k$& the number of agents.    \\
\hline
$C_i$& the positive cascade generated by the $i$-th agent.    \\
\hline
$a_i$& the seed set of cascade $C_i$.    \\
\hline
$B_i$& the budget of the seed set of cascade $C_i$.    \\
\hline
$\overline{\gamma}()$& social utility of the game.    \\
\hline
$\overline{\delta}_i()$& private utility of the $i$-th agent.    \\
\hline
$\overline{\sigma}_i(S)$& the expected number of $C_i$-active nodes under strategy $S$.    \\
\hline
$t_A^g(u)$& the activation time of node $u$ in $g$ under the full-action $A$.    \\
\hline
\end{tabular}}
\caption{\textbf{Notations.} }
\label{table:notations}
\end{table}

\subsubsection{Multiple Cascade}
Suppose there are multiple cascades each of which is generated by its own seed set. We denote by $C_r$ the cascade generated by rumor with a fixed seed set $a_r$. The basic definitions are shown as follows.

\begin{definition}
For a certain cascade $C$, we call a node $C$-\textit{active} (resp. $\overline{C}$-\textit{active}) if it is activated (resp. not activated) by cascade $C$. 
\end{definition}
\begin{definition}[Cascade Priority]
Each cascade is assigned a distinct priority and we assume that the rumor always has the highest priority. We denote by $\mathrm{Priority}(C)$ the priority of cascade $C$ and, for two cascades $C_1$ and $C_2$,  $\mathrm{Priority}(C_1) < \mathrm{Priority}(C_2)$ if  and only if cascade $C_2$ has a higher priority than that of cascade $C_1$. 
\end{definition}

%
\begin{definition}[Activation Order]
\label{def:activation order}
Let $b_u^t$ be the set of the node $v$ such that $v \in N_u$ and $u$ has not tried to activate $v$ before time $t$. At time step $t$, an active node $u$ will uniformly at random select a node in $b_u^t$ to activate\footnote{Other kinds of activation orders will be discussed later in Sec. \ref{sec:further}.}. 
\end{definition}

Recall that the PIC model represents the private communication and consequently a user cannot know whether the other users have been activated or not. Therefore, one user may attempt to activate another user who has already been activated by others. 

\subsubsection{Diffusion Process}
Given a PIC network $G$ and the seed sets of the cascades, the diffusion process unfolds in discrete, as described in following.
\begin{itemize}
\item Time step $0$. Each cascade $C$ activates its seed nodes. If one node is selected by more than one cascades, it will be activated by the cascade with the highest priority.

\item Time step $t>0$. Each active node $u$ randomly select one node $v$ in $b_u^t$ and activates $v$ with a success probability of $p_{(u,v)}^G$, where each node in $b_u^t$ has the same probability to be selected by $u$. If $u$ is $C$-active and $u$ successfully activates $v$ then $v$ becomes $C$-active. If a node is successfully activated by two or more neighbors pertaining to different cascades, it will be activated by the cascade with the highest priority. 
\end{itemize}

The PIC model is a probabilistic model where the randomness comes from that (a) at each step who to select to activate and (b) whether the activation succeeds. The following definition shows a derandomization of the diffusion process under the PIC model.

\begin{algorithm}[t]
\caption{ \textbf{Realization Generation}}\label{alg:realization}
\begin{algorithmic}[1]
\State \textbf{Input}: A PIC network $G=(V,E)$.
\State \textbf{Output}: \small{A realization $g=(V_g,E_g)$ together with $\alpha_u^g$ for each node $u$ and $p_e^g$ for each $e$.}
\State $V_g \leftarrow V$ and $E_g \leftarrow E$; 
\For {each edge $e \in E$}
		\State $rand \leftarrow$ a random number from 0 to 1 generated in uniform; 
		\If {$rand  \leq p_{e}^G$}
		\State $p_{e}^g \leftarrow 1$;
		\Else
		\State $p_{e}^g \leftarrow 0$;
		\EndIf 
\EndFor
\For {each node $u \in V$}
		\State $\alpha_u^g \leftarrow$ a permutation of $N_u$ generated uniformly at random; 
\EndFor
\State Return $g$, $p_e^g$ and $\alpha_u^g$;
\end{algorithmic}
\end{algorithm}

\begin{definition}[Realization]
\label{def:realization}
A \textit{realization} \cite{tong2015adaptive} $g=(V_g,E_g)$ of a PIC network $G=(V,E)$ is a special PIC network randomly constructed by Algorithm \ref{alg:realization}. First, $V_g=V$ and $E_g=E$. The propagation probability $p_e^g$ of each edge $e$ in $g$ is either 0 or 1 determined in random. In particular, for each edge $e$, the probability that $p_e^g=1$ (resp. $p_e^g=0$) is $p_e^G$ (resp. $1-p_e^G$). Each node $u$ randomly decides a permutation $\alpha_u^g$ (i.e., an order) of all its out-neighbors $N_u$ in $G$ where each possible permutation of $N_u$ has the same probability to be selected by $u$. We take a permutation $\alpha_u^g$ as a one-to-one mapping from $N_u$ to $\{1,...,|N_u|\}$. In $g$, the activation order of the our-neighbors of $u$ is determined by the permutation $\alpha_u^g$. That is, when $u$ becomes active, $u$ selects its neighbor to activate one by one according to the order given by  $\alpha_u^g$. The cascade priority in $g$ remains the same as that in $G$. Furthermore, we assign a weight of each edge in $g$. Suppose $u$ has $d_u$ out-neighbors $v_1,...,v_{d_u}$ in $G$. For $1\leq i \leq d_u$, the weight $w^g{(u,v_i)}$ of edge $(u,v_i)$ is $j$ in $g$ if $\alpha_u^g(v_i)=j$. For two nodes $u$ and $v$, let $\mathrm{dis}^g(u,v)$ be the length of the shortest path from $u$ to $v$ in $g$. For a node set $V^{'}$ and a node $v$, define that $\mathrm{dis}^g(V^{'},v)=\mathrm{min}_{u \in V^{'}}\mathrm{dis}^g(u,v)$. For a certain realization $g$, let $\mathrm{Pr}[g]$ be the probability that $g$ can be generated by Algorithm \ref{alg:realization}. Let $\mathcal{G}$ be the set of all possible realizations.
\end{definition}
One can see that each realization $g$ corresponds to a basic event of the PIC model. If an edge $(u,v)$ has a probability of 1 in $g$, then it means $u$ can successfully activate $v$. The weight $w^g{(u,v)}$ of an edge $(u,v)$ implies that if $u$ is activated at time $t$ then it will try to activate $v$ at time $t+w^g{(u,v)}$. The following theorem shows the relationship between a PIC network and its realizations.

\begin{theorem}
\label{theorem:equivalent}
Given the seed set of each cascade, the following two diffusion processes are the equivalent to each other, with respect to the distribution of the spreading results. 
\begin{itemize}
\item a. Execute the stochastic diffusion process on the PIC network $G$.
\item b. Randomly generate a realization $g$ of $G$ according to Algorithm \ref{alg:realization}, and execute the deterministic diffusion process on $g$. 
\end{itemize}
\end{theorem} 
\begin{proof}
See appendix \ref{sec:proof_theorem_equivalent}.
\end{proof}

In the next section, we will discuss the property of the rumor blocking game where Theorem \ref{theorem:equivalent} plays an important role.



\section{A Game-theoretical analysis}
\label{sec: game}
We assume each cascade is generated by an \textit{agent} who decides the seed set of that cascade. For example, an agent can be a company that posts an  advertisement for its product. In the traditional rumor blocking problem, it is assumed that there is an authority who generates a single positive cascade. However, the real social networks are extremely large and such an authority is not efficient and sometimes even unfeasible. In this section, we consider the scenario that there are multiple positive cascades generated by different agents and each agent aims to limit the spread of rumor by itself, which forms a game between the agents. 

\subsection{Some Notations}

Suppose there are $k$ positive cascades $\{C_1,...,C_k\}$ generated by $k$ agents, respectively. Together with the rumor $C_r$ there are totally $k+1$ cascades in the network. 

\begin{definition}[Action Space]
\label{def: action_space}
Associated with each agent, there is an action space which is a collection of seed sets that they can select. We denote by $\mathcal{A}_i$ the action space of the $i$-th agent. 
\end{definition}

$\mathcal{A}_i$ is usually not equal to $2^{V}$. For example, a company can only convince the users who like the product to be the seed users. The most considered constraint is the budget constraint where each agent can select at most a certain number of seed nodes.



\begin{definition}[Full-action]
\label{def: full-action}
A \textit{full-action} $A=(a_1,...,a_k) \in \mathcal{A}_1 \times \mathcal{A}_2 \times ... \times \mathcal{A}_k$ specifies the seed sets selected by the agents. 
\end{definition}

Instead of taking a single action from the action space, an agent may decide an action according to a distribution $s$ over all of their actions. We called such a distribution $s$ as a strategy and denote by $\mathcal{S}_i$ the set of all strategies of the $i$-th agent. 

\begin{definition}[Strategy Space]
The \textit{strategy space} $\mathcal{S}_i$ of the $i$-th agent is a set of the distributions over the actions in $\mathcal{A}_i$. For each $s \in \mathcal{S}_i$ and $a \in \mathcal{A}_i$, we use $\mathrm{Pr} [a|s]$ to denote the probability that action $a$ is taken under the strategy $s$. We denote by $\emptyset$ the empty strategy where $\mathrm{Pr} [a|\emptyset]=0$ for each action $a$.
\end{definition}

In analogy with Def. \ref{def: full-action}, we have the following term for strategies.

\begin{definition}[Full-strategy]
A \textit{full-strategy} $S=(s_1,...,s_k)\in \mathcal{S}_1 \times \mathcal{S}_2 \times ... \times \mathcal{S}_k$ specifies the strategy adopted by each agent, where $s_i$ is the strategy adopted by the $i$-th agent. For a full-strategy $S$ and a full-action $A$, let $\mathrm{Pr} [A|S]$ be the probability that $A$ is implemented under $S$.
\end{definition}

\subsection{Social Utility}
\label{sec: social}

For the rumor blocking game, the social utility is the number of the users that are not activated by rumor. 
\begin{definition}[Social Utility]
For a full-strategy $S$ of the agents, we use $\overline{\gamma}(S)$ to denote the expected number of $\overline{C}_r$-active nodes. 
\end{definition}

We are particularly interested in the marginal return of $\overline{\gamma}(S)$ resulted by adding more agents to game. For the purpose of analysis, we introduce the following notations.

\begin{definition}
For a full-strategy $S=(s_1,...,s_k)$, a full-action $A=(a_1,...,a_k)$ and an integer $i \leq k$, let $S^i=(s_1,...,s_i,\emptyset,...,\emptyset)$ and $A^i=(a_1,...,a_i,\emptyset,...,\emptyset)$. For a full-action $A=(a_1,...,a_k)$, we denote by $A(i,a_i^{'})$ the full-action where the $i$-th agent replaces its action $a_i$ in $A$ by $a_i^{'}$. Similarly we have the notation $S(i,s_i^{'})$ for a full-strategy $S$ and a strategy $s_i^{'}$ of the $i$-th agent. 
\end{definition}

Intuitively, for $i \leq j$, $\overline{\gamma}(S^{i}(j,s))-\overline{\gamma}(S^{i})$ denotes the marginal return when the $j$-th agent join the game with a strategy $s$. The following result indicates that the social utility of our rumor blocking game has the property of diminishing marginal return.

\begin{theorem}
\label{theorem: submodular}
$\overline{\gamma}(S^{i_1}(i_3,s^{*}))-\overline{\gamma}(S^{i_1}) \geq \overline{\gamma}(S^{i_2}(i_3,s^{*}))-\overline{\gamma}(S^{i_2})$, for $1 \leq i_1 \leq i_2 \leq i_3 \leq k$ and any strategy $s^{*} \in \mathcal{S}_{i_3}$.
\end{theorem}

\textbf{The Proof of Theorem \ref{theorem: submodular}.} In the rest of this section, we provide a sketch of the proof of Theorem \ref{theorem: submodular}. The details can be found in the appendix. 

For a full-action $A=(a_1,...,a_k)$, let $\gamma(A)$ be the expected number of $\overline{C}_r$-active nodes under $A$. For a full-action $A=(a_1,...,a_k)$ and a realization $g$, let  $\gamma^g(A)$ be the number of $\overline{C}_r$-active nodes in $g$ under $A$. By Theorem \ref{theorem:equivalent},
\begin{equation*}
\gamma(A)=\sum_{g \in \mathcal{G}} \mathrm{Pr}[g] \cdot \gamma^g(A)
\end{equation*}

The key to proving Theorem \ref{theorem: submodular} is that $\gamma(A)$ only depends on the union of the actions in $A$, shown as follows.

\begin{lemma}
\label{lemma:set_function}
For a full-action $A=(a_1,...,a_k)$, let $A^{*}=a_1 \cup a_2 \cup ... \cup a_k$ be the union of the seed sets of the agents. $\gamma(A)$ is a set function on $A^{*}$. That is, for any two full actions $A_1$ and $A_2$, $\gamma(A_1)=\gamma(A_2)$ if $A_1^{*}=A_2^{*}$.
\end{lemma}
\begin{proof}
See Appendix \ref{sec:proof_lemma_set_function}.
\end{proof}
It is worthy to note that in some other models $\gamma()$ may not be a set function of the union of positive seed sets, as discussed in Sec. \ref{sec:further}. Since $\gamma()$ is a set function, for any $X \subseteq V$, let $\gamma^g(X)$ be the number of $\overline{C}_r$-active nodes in $g$ when the union of the seed sets of positive cascades is $X$. 

A set function $f : 2^{V} \rightarrow \mathbb{R}$ is called monotone increasing if $f(X) \leq f(Y)$ for any $X \subseteq Y$. A set function $f : 2^{V} \rightarrow \mathbb{R}$ is called \textit{submodular} if $f(X)+f(Y)\geq f(X \cap Y)+f(X \cup Y)$ for any $X$ and $Y \in 2^{V}$.
\begin{lemma}
\label{lemma:submodular}
$\gamma()$ is monotone increasing and submodular.
\end{lemma}
\begin{proof}
See Appendix \ref{sec:proof_lemma_submodular}.
\end{proof}

By Lemma  \ref{lemma:submodular}, $\gamma(A^{i_1}(i_3,a))-\gamma(A^{i_1}) \geq \gamma(A^{i_2}(i_3,a))-\gamma(A^{i_2})$, for any full-action $A$, $1 \leq i_i \leq i_2 \leq i_3 \leq k$ and $a \in \mathcal{A}_{i_3}$. Since $\overline{\gamma}(S)=\sum_A \Pr[A|S] \cdot \gamma(A)$, Theorem \ref{theorem: submodular} follows immediately.

\subsection{Private Utility and the Nash Equilibrium}
\label{sec:rumor_aware}
Now let us consider the private utility of the games. We consider two games depending on whether or not the agents are able to distinguish rumor from other positive cascades.

\subsubsection{Rumor-aware Game}
\label{subsec:ra-game}
In a social network, the agents are able to identify the rumor when content of rumor is completely different from the facts. Assuming the agents are aware of the rumor, the private utility $\overline{\delta}_i()$ of the $i$-th agent is 
\begin{equation}
\label{eq:private_1}
\overline{\delta}_i(S)=\overline{\gamma}(S)-\overline{\gamma}(S(i,\emptyset)),
\end{equation}
which is the effort made by the $i$-th agent to limit the spread of rumor. For a full-action $A$, we use $\delta_i(A)=\gamma(A)-\gamma(A(i,\emptyset))$ to denote the private utility of the $i$-th agent under $A$. We term this game as the \textit{rumor-aware game}.

Since the agents aim to maximize $\overline{\delta}_i()$ and the rumor has the highest priority, we can assume $a_i \cap a_r =\emptyset$ for any $a_i \in \mathcal{A}_i$ without loss of generality.

An agent may change their strategy to gain more private utilities according to the strategies of other agents. For a full-strategy $S$, it reaches the Nash equilibrium if no player can gain more utility by changing their own strategy. That is,
\begin{equation*}
\overline{\delta}_i(S) \geq \overline{\delta}_i(S(i,s_i^{'}))
\end{equation*}
for each $i$, $1 \leq i \leq k$, and each $s_i^{'}$ in $\mathcal{S}_i$. Due to Nash \cite{nash1951non}, the finite k-agent non-cooperative game always has at least one Nash equilibrium. In the following, we will show that any Nash equilibrium of the rumor-aware game guarantees the social utility with a provable ratio compared to the optimal strategy. 

A game is a \textit{valid utility system} if, under any full-strategy $S$, (a) the private utility is not less than the marginal social utility and (b) the total private utility is not larger than the social utility. That is 
\begin{equation}
\label{eq:utility}
\overline{\delta}_i(S) \geq \overline{\gamma}(S)-\overline{\gamma}(S(i,\emptyset))
\end{equation}
for each $i$, and
\begin{equation}
\label{eq:valid}
\sum_{i=1}^{k} \overline{\delta}_i(S) \leq \overline{\gamma}(S)
\end{equation}

\begin{theorem}
\label{theorem:rumor_aware_valid}
The rumor-aware game is a valid utility system under the PIC model.
\end{theorem}
\begin{proof}
See Appendix \ref{sec:proof_theorem_rumor_aware_valid}.
\end{proof}

According to Vetta \cite{vetta2002nash}, if the social utility is a submodular set function, the Nash equilibrium of the game guarantees the social utility by a factor of 2, and therefore we have the following result immediately.

\begin{corollary}
\label{corollary:2-approximate}
Suppose a full-strategy $S$ forms a Nash equilibrium of the rumor-aware game, and let $\Omega$ be the full-strategy such that $\overline{\gamma}(\Omega)$ is maximized. Then, $\overline{\gamma}(S) \geq  (\frac{1}{2}) \cdot \overline{\gamma}(\Omega)$.
\end{corollary}

Recall that in the rumor-aware game the social utility is the expected number of $\overline{C}_r$-active nodes while the private utility is the effort made by each agent to limit the spread of rumor. Due to the nature of non-cooperation, each agent only concerns their marginal contribution. Nevertheless, Corollary \ref{corollary:2-approximate} shows that the social utility will not be arbitrarily far from the optimal and in fact it guarantees a 2-approximation. Intuitively speaking, even if there is no powerful authority dealing with rumor in a social network, the rumor can be efficiently blocked by the users who participate and propagate positive information.

A \textit{pure} full-strategy is a special full-strategy where each the $i$-th agent decides to carry out one specific action. When a full-strategy $S=(s_1,...,s_k)$ is pure, each $s_i$ is a vector of 0 and 1. In other words, the pure full-strategy reduces to the full-action. The Nash equilibrium formed by a pure full-strategy is called pure Nash equilibrium. Note that there always exists an optimal strategy $\Omega$ which is pure. For general games, the Nash equilibrium may not be pure. However, the agents usually make pure strategies instead of making decisions according to a distribution. The following result shows that for the pure Nash equilibrium always exists in the rumor-aware game. 

\begin{theorem}
\label{theorem:exists_full}
For the rumor-aware game, there exists a full-action $\Phi$ such that $\delta_i(\Phi) \geq \delta_i(\Phi(i,a_i^{'}))$ for each $1 \leq i \leq k$ and each action $a_i^{'}$ of the $i$-th agent.
\end{theorem}

\begin{proof}
Let $\Phi_0$ be an arbitrary full-action and consider the following process to generate a series of full-actions $\Phi_0,...,\Phi_m$. For the full-action $\Phi_j$, if for some $i$  there exists an action $a_i \in \mathcal{A}_i$ such that  $\delta_i(\Phi_j) < \delta_i(\Phi_j(i,a_i))$, we denote $\Phi(i,a_i)$ as $\Phi_{j+1}$. By such process, we finally obtain a sequence of full-actions $\Phi_j$. According to the construction, for any $\Phi_j$ and $\Phi_{j+1}$, there exists some $i$ such that $\delta_i(\Phi_j) < \delta_i(\Phi_{j+1})$, which implies $\gamma(\Phi_j) < \gamma(\Phi_{j+1})$. Therefore, any two full-actions in the sequence cannot be identical. Since the action space is finite, the sequence $\Phi_j$ must be finite and the last full-action reaches the pure Nash equilibrium.
\end{proof}

The proof of Theorem \ref{theorem:exists_full} implies that we can build a Nash equilibrium starting from any full-action by increasing the private utility of some agent.

\textbf{The Simple Game.} In the prior works, the goal is to block the rumor by introducing a competing cascade. Due to the expense of activating seed nodes, there is a budget of the seed nodes. In our rumor blocking game, such a budget is distributed to multiple agents. Suppose the budget is distributed to $k$ agents where each agent has one budget, the agents make decisions in turn, and, the agents always make pure decisions. The evolution of the Nash equilibrium is shown in Algorithm \ref{alg:first_equilibrium}. We denote the game under such a setting as \textit{Simple Game}. 

\begin{algorithm}[t]
\caption{ \textbf{Simple Game}}\label{alg:first_equilibrium}
\begin{algorithmic}[1]
\State $A \leftarrow (\emptyset,...,\emptyset)$; 
\State $\text{sign} \leftarrow true$; 
\While {$A^{'} \neq A$}
	\State $A \leftarrow A^{'} $;
	\For {$i=1:k$}
		\State $v \leftarrow \argmax_{v \in V} \delta_i(A(i,\{v\}));$
		\State $A^{'} \leftarrow A(i,\{v\})$;
	\EndFor
\EndWhile
\State Return $A$;
\end{algorithmic}
\end{algorithm}




\textbf{Under the budget constraint.} The result shown in Corollary \ref{corollary:2-approximate} requires that each agent follows the best response policy. However, the agent in real case may not be able to efficiently\footnote{Ideally decisions should be made in polynomial time.} find such an optimal action that maximizes the private utility. Under the budget constraint, each agent can select at most a certain number of seed nodes.  In this case, finding the bests response is NP-hard \cite{budak2011limiting} so the polynomial approximation response is the best that each agent can adopt. According to Lemma \ref{lemma:submodular}, given the seed sets of other agents, $\delta_i()$ is also monotone increasing and submodular, and therefore the $i$-th agent can easily obtain an action $a_i$ such that $\gamma(A(i,a_i)) \geq (1-e^{-1}) \cdot \gamma(A(i,a_i^{*})$, where $a_i^{*}$ is the best response \cite{nemhauser1978analysis}. If all the agents adopt the approximation action, the game finally reaches an approximate Nash equilibrium. The next result shows that such an equilibrium guarantees the social utility within a factor of $\frac{2e+1}{e+1}$. 

\begin{lemma}
\label{lemma:pre_second}
Let $A=(a_1,...,a_k)$ be a pure Nash equilibrium under the approximate response. Then $\gamma(A^{*} \cup \Omega^{*}) \leq \frac{2e+1}{e+1}\cdot \gamma(A^{*})$ where $\Omega$ is the optimal pure full-action that maximizes $\gamma()$.
\end{lemma}
\begin{proof}
See Appendix \ref{sec:proof_lemma_pre_second}.
\end{proof}

\begin{theorem}
\label{theorem:second}
If each agent adopts the $(1-e^{-1})$-approximate response, the Nash equilibrium guarantees an $\frac{2 \cdot e-1}{e-1}$-approximation with respect of the expected number of $\overline{C}_r$-active nodes. 
\end{theorem}
\begin{proof}
Since $\gamma()$ is a set function and $\gamma(\Omega^{*}) \leq \gamma(A^{*} \cup \Omega^{*})$, the theorem directly follows from Lemma \ref{lemma:pre_second}. 
\end{proof}

\subsubsection{Rumor-oblivious Game}
\label{sec:ro-game}
In another issue, the rumor may be well disguised such that they cannot be distinguished from the genuine news. In this case, the best that an agent can do is to maximize the spread of its own cascade. Therefore, the private utility $\overline{\delta}_i()$ of the $i$-th agent is $\overline{\delta}_i(S)=\overline{\sigma}_i(S)$, where
\begin{equation}
\label{eq:private}
\overline{\sigma}_i(S)=\sum_{A} \mathrm{Pr} [A|S] \cdot \sigma_i(A)
\end{equation}
is the expected number of $C_i$-active nodes under $S$ and $\sigma_i(A)$ is the expected number of $C_i$-active nodes under $A$. Such a game is called rumor-oblivious game. In the following, we will show that the rumor-oblivious game also forms a valid utility system. However, the proof slightly differs from that of the rumor-aware game.

\begin{lemma}
\label{lemma:rumor_oblivious_valid_utility}
The rumor-oblivious game is a valid utility system under the PIC model.
\end{lemma}
\begin{proof}
See Appendix \ref{sec:proof_lemma:rumor_oblivious_valid_utility}.
\end{proof}

We have the following result due to Vetta \cite{vetta2002nash}
\begin{corollary}
For any Nash equilibrium $S$ of the rumor-oblivious game, $\overline{\gamma}(S) \geq  (\frac{1}{2}) \cdot \overline{\gamma}(\Omega)$, where $\Omega$ is the full-strategy maximizing $\overline{\gamma}()$.
\end{corollary}

As  discussed in prior works e.g. \cite{bharathi2007competitive}, given the actions of other agents, $\sigma_i(A)$ is also monotone and submodular with respect to the seed set of the $i$-th agent. Therefore, similar to the analysis in Sec. \ref{sec:rumor_aware}, the agents in the rumor-oblivious game are also able to make the $(1-1/e)$-approximate pure response. However, unlike the rumor-aware game, there may not be a $\frac{2 \cdot e-1}{e-1}$-approximation equilibrium for the rumor-oblivious game. This is because the pure Nash equilibrium may not exist in the rumor-oblivious game.

\section{Discussions on \\Cascade Priority and Activation Order}
\label{sec:further}
In this section, we provide several observations concerning the competitive diffusion model. The discussion herein may help us further understand the scenario when more than two cascades exist. We introduce this following two types of cascade priority.

\begin{definition}[Homogeneous \& Heterogeneous Cascade Priority]
The cascade priority is homogeneous if the priority of the cascades is the same for each user. Otherwise, it is called heterogeneous cascade priority.
\end{definition}


\begin{figure}[t]
\captionsetup{justification=centering}
\subfloat[Action $A_1$.]{\label{fig:further_a1}\includegraphics[width=0.18\textwidth]{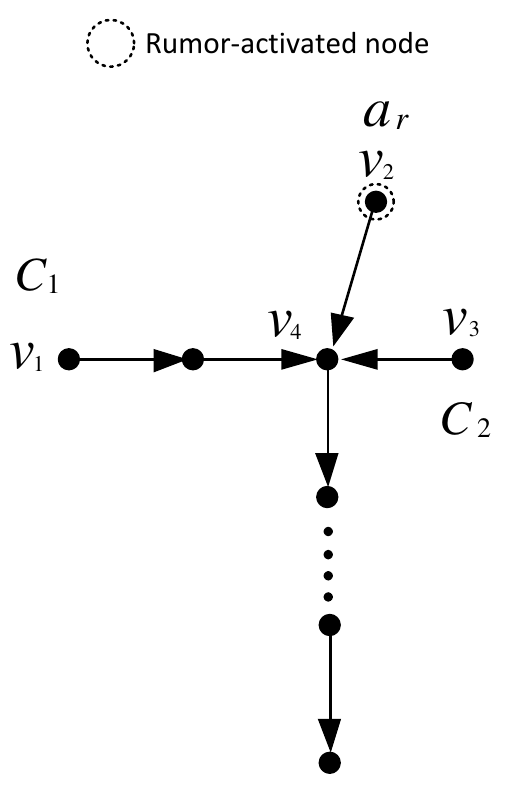}} \hspace{15mm}
\subfloat[Action $A_2$.]{\label{fig:further_a2}\includegraphics[width=0.18\textwidth]{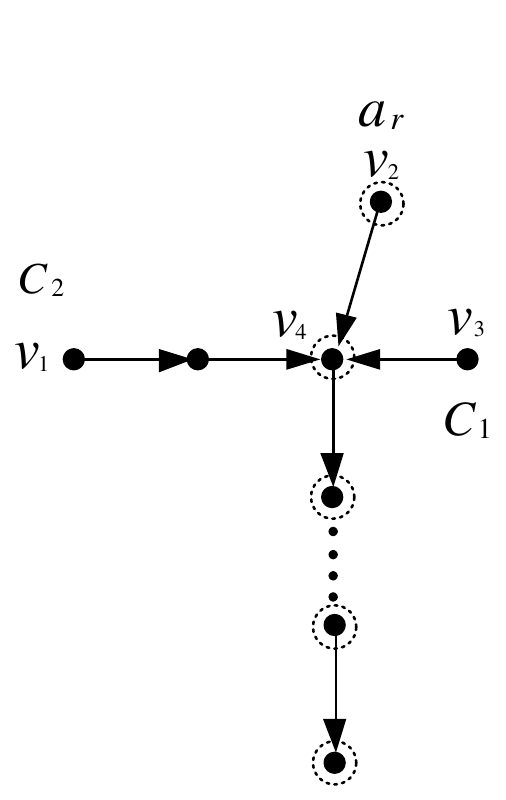}} \hspace{0mm}

\subfloat[Second example.]{\label{fig:further_b}\includegraphics[width=0.18\textwidth]{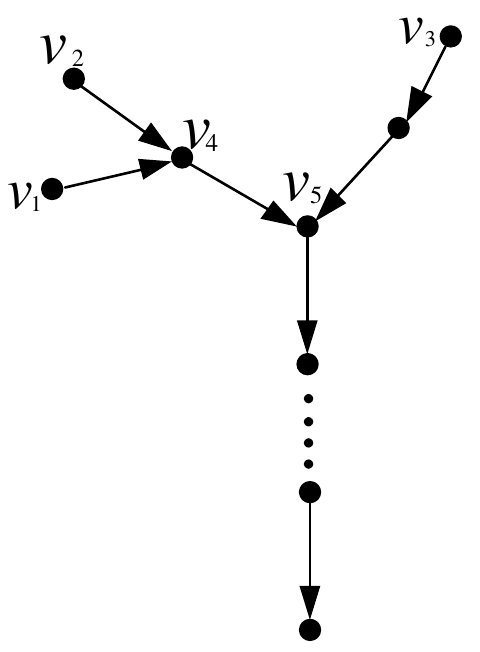}} \hspace{15mm} 
\subfloat[Third example.]{\label{fig:further_c}\includegraphics[width=0.18\textwidth]{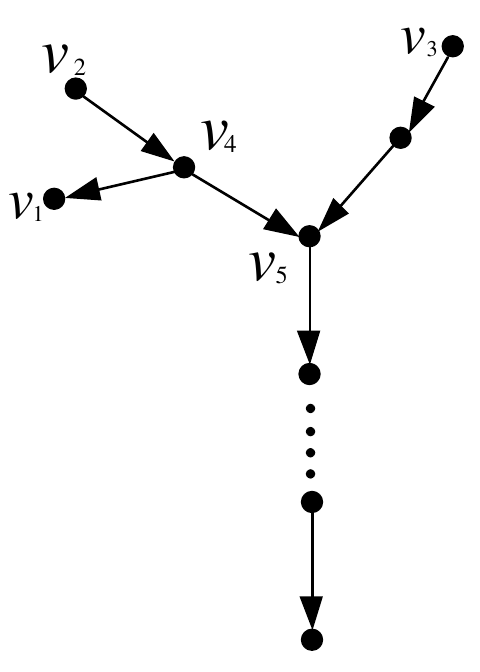}} \hspace{0mm}
\caption{\small \textbf{Examples.}} 
\label{fig:g_vs_f}
\vspace{-3mm} \normalsize
\end{figure}

We list some observations, as follows.

\begin{fact}
Under the homogeneous cascade priority, if rumor does not have the highest priority, $\gamma(A)$ is not a set function anymore.
\end{fact}
\begin{proof}[An example.]
Consider the network shown in Fig. \ref{fig:further_a1} and \ref{fig:further_a2}, where $p_e^G=1$ for each edge $e$. Suppose $a_r=\{v_2\}$ and there are two agents with cascades $C_1$ and $C_2$, respectively. Suppose that $\mathrm{Priority}(C_1) \leq \mathrm{Priority}(C_r) \leq \mathrm{Priority}(C_2)$. One can easily check that action $A_1=(\{v_1\},\{v_3\})$ and $A_2=(\{v_3\},\{v_1\})$ result different values of social utility. Under action $A_1$, $v_4$ will be activated by the rumor seed $v_2$ and therefore many nodes will be later activated by rumor spreading from $v_4$. However, under $A_2$, only the node $v_2$ will be activated by rumor as $v_4$ will be activated by cascade $C_1$. Thus, $\gamma(A_1) \neq \gamma(A_2)$ even if $A_1^{*}=A_2^{*}$.
\end{proof}

\begin{fact}
Under the heterogeneous cascade priority, the social utility $\gamma(A)$ is not monotone increasing.
\end{fact}
\begin{proof}[An example.]
The heterogeneous cascade priority setting has been adopted in prior works e.g. \cite{borodin2010threshold}. We observe that under this setting, the social utility may decrease when more agents join in game. Consider the illustrative example shown in Fig. \ref{fig:further_b}, where each $p_e^G=1$ for each edge $e$ and $a_r =\{v_3\}$. Suppose there are two agents $C_1$ and $C_2$, and for $v_4$ and $v_5$, the priority of the cascades is $$\mathrm{Priority}(C_2) < \mathrm{Priority}(C_1) < \mathrm{Priority}(C_r)$$ and $$\mathrm{Priority}(C_1) \leq \mathrm{Priority}(C_r) \leq \mathrm{Priority}(C_2),$$ respectively. Consider two actions $A_1$ and $A_2$ where $A_1=(\{v_1\},\emptyset)$ and $A_2=(\{v_1\},\{v_2\})$. Under $A_1$ there is only one agent in the game and $v_5$ will be activated by $C_1$ because $C_1$ has the higher priority than $C_r$ at $v_5$. However, when another agent joins the game as shown by $A_2$, $v_5$ will become $C_r$-active because $v_4$ will be activated by $C_2$ and $\mathrm{Priority}(C_r) \leq \mathrm{Priority}(C_2)$ at $v_4$. Therefore, under this setting, $\gamma(A)$ may not be monotone increasing with respect to $A^{*}$.
\end{proof}

\begin{fact}
Under the homogeneous cascade priority, if an active user only attempts to activate inactive neighbors, the social utility $\gamma(A)$ is not monotone increasing.
\end{fact}

\begin{proof}[An example.]
Note that we in this paper assume that a node may try to activate the neighbor that has been active. It is worthy to note that if each node only attempts to activate inactive neighbors, then $\gamma(A)$ is not monotone increasing under the PIC model. An example is shown in Fig. \ref{fig:further_c}. Again we assume that each edge has the probability of $1$, $a_r=\{v_2\}$ and there are two agents. Suppose each node only selects inactive node to activate and $v_4$ will activate $v_1$ and $v_5$ in order after becoming active. Consider the two actions $A_1=(\{v_3\}, \emptyset)$ and $A_2=(\{v_3\},\{v_1\})$. One can see that $v_5$ will be activated by $C_1$ under $A_1$ because at the second step $v_4$ will activate $v_1$ after rumored by $v_2$. Nevertheless, if another agent participates and selects $v_1$ as the seed node, as shown in $A_2$, then $v_5$ will be rumor-activated by $v_4$, because at the second step $v_4$ will not try to activate $v_1$ as $v_1$ has been activated in the first step. Thus, when more agents come to limit the spread of rumor, the rumor may surprisingly spread more widely.
\end{proof}

As shown above, under certain settings the model does not have good properties anymore and consequently, the rumor blocking problem becomes more complicated in such scenarios.

\begin{figure*}[t]
\centering
\subfloat[Facebook with $p_{(u,v)}^G=0.01$]{\label{fig:facebook_001_1_30}\includegraphics[width=0.24\textwidth]{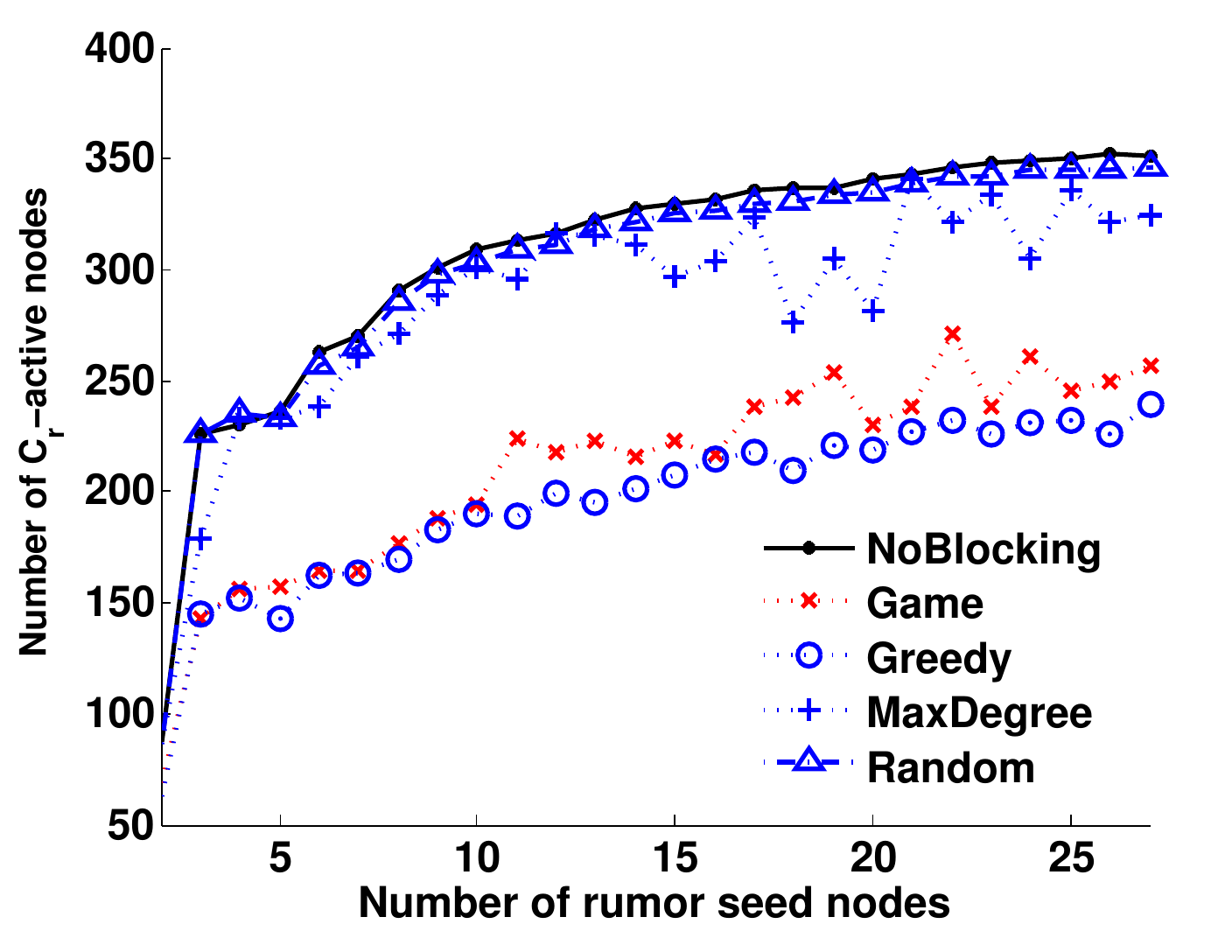}} \hspace{2mm} 
\subfloat[Facebook with $p_{(u,v)}^G=0.1$]{\label{fig:facebook_01_1_30}\includegraphics[width=0.24\textwidth]{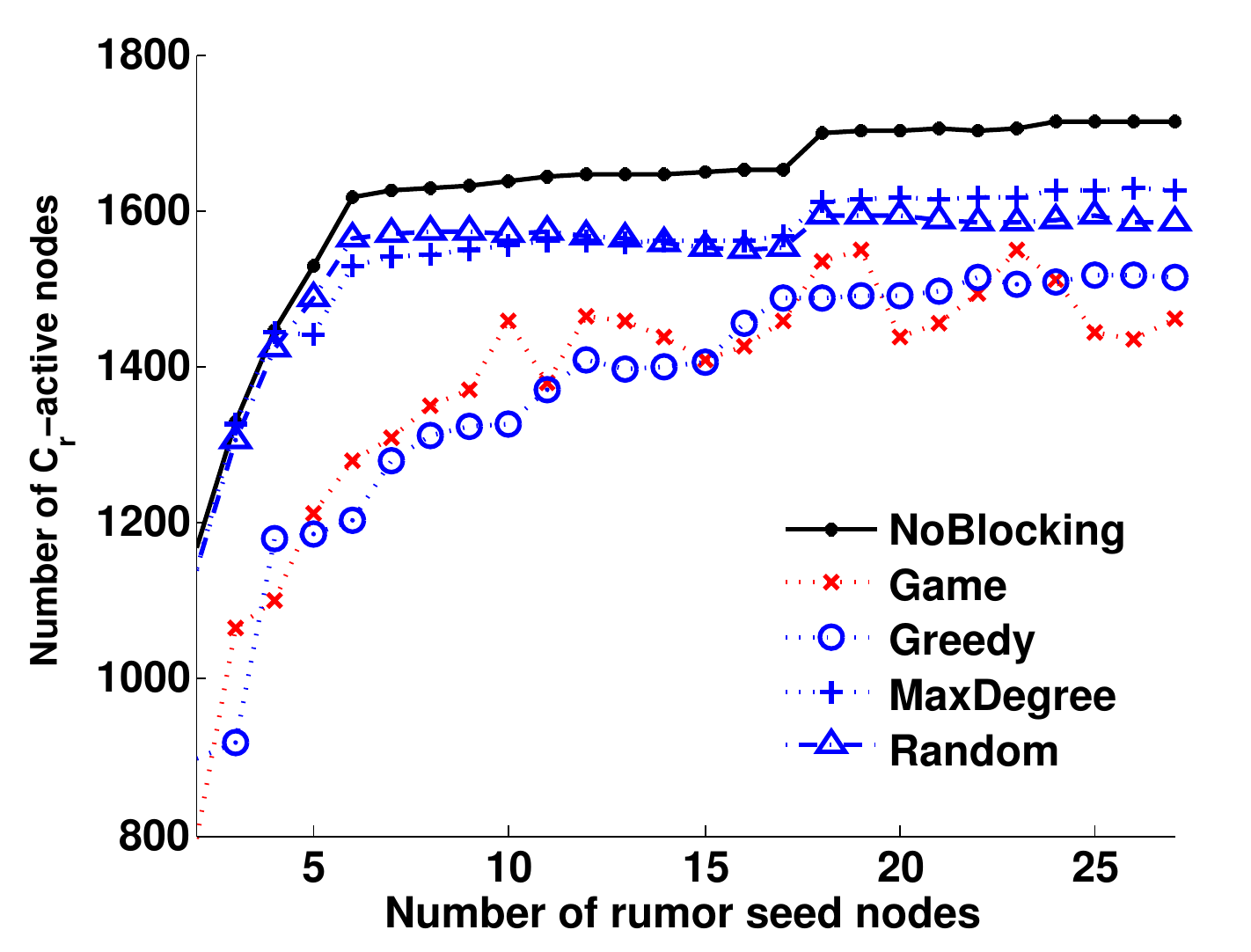}} \hspace{2mm} 
\subfloat[Facebook under weighted cascade setting]{\label{fig:facebook_cas_1_30}\includegraphics[width=0.24\textwidth]{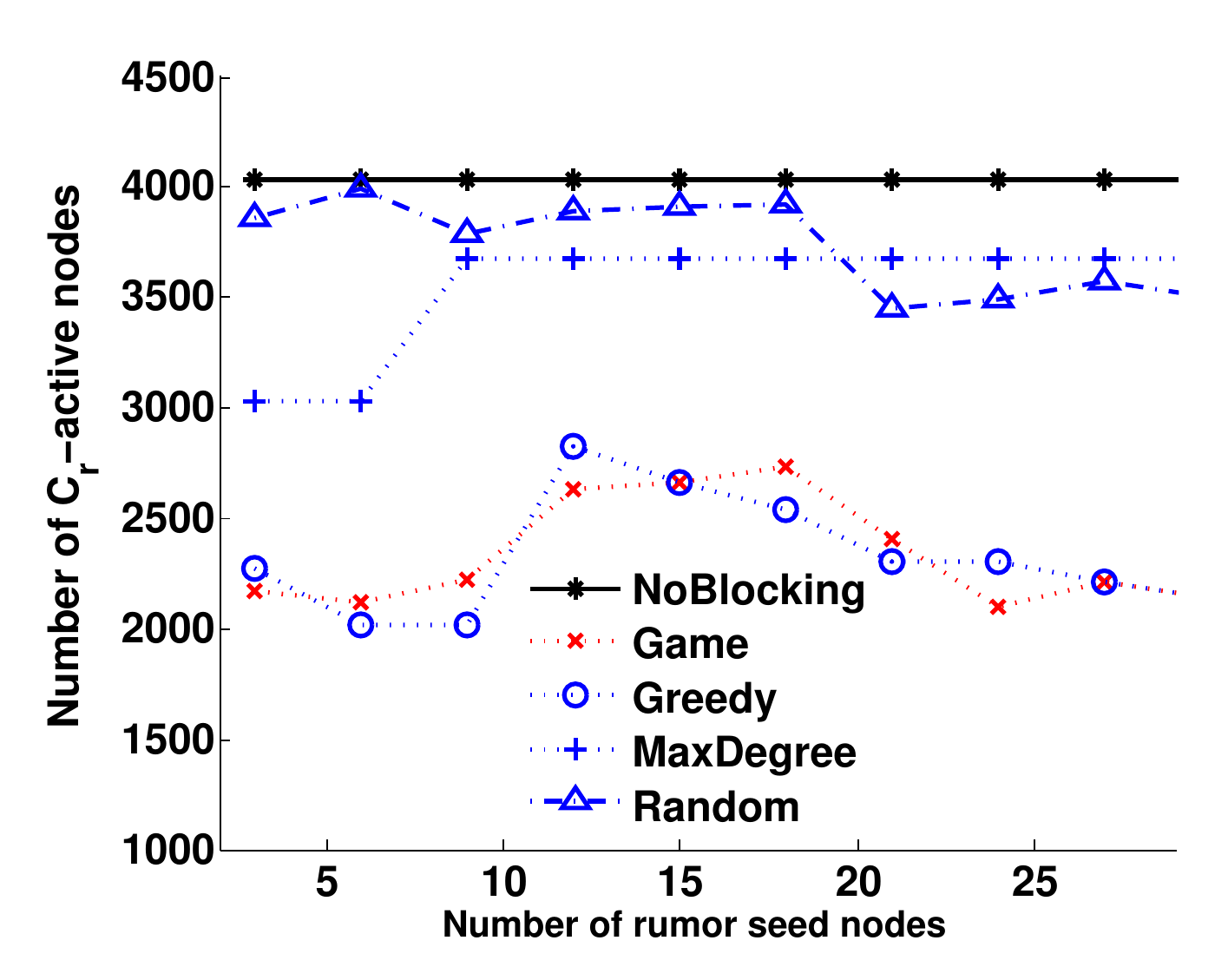}} \hspace{0mm} 
\subfloat[Hep with $p_{(u,v)}^G=0.1$]{\label{fig:hep_01_1_30}\includegraphics[width=0.24\textwidth]{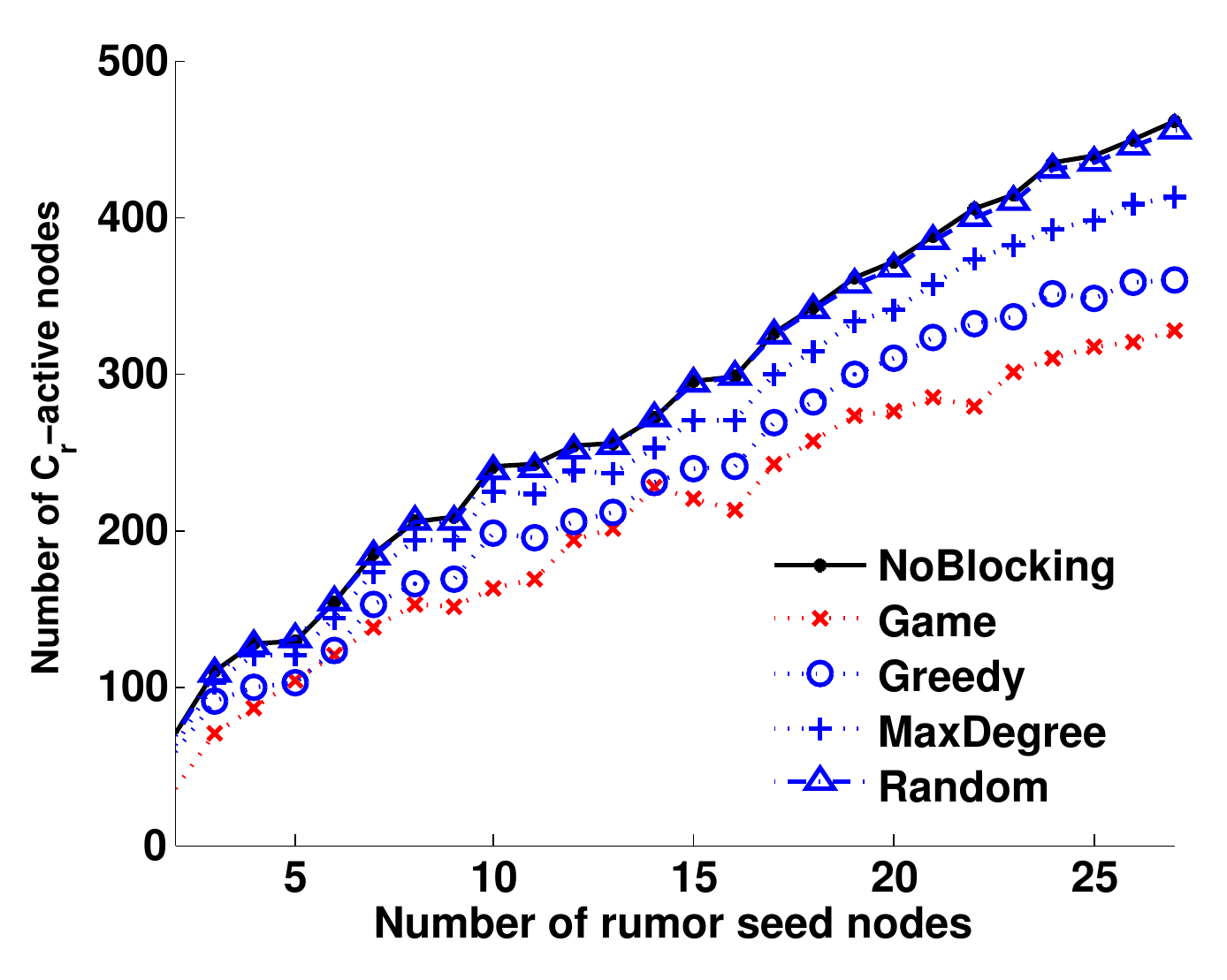}} \hspace{0mm} 



\caption{Results of the first experiment. The y-axis and x-axis denote the expected number of $C_r$-active nodes and the number of rumor seed nodes, respectively. Each graph gives
five curves plotting the number of $C_r$-active nodes under NoBlocking, Game, Greedy, MaxDegree, and Random, respectively. } 
\label{fig:exp1}
\end{figure*}

\begin{figure*}[t]
\centering
\subfloat[Facebook with $p_{(u,v)}^G=0.1$ and $|a_r|=20$]{\label{fig:facebook_01_r20}\includegraphics[width=0.32\textwidth]{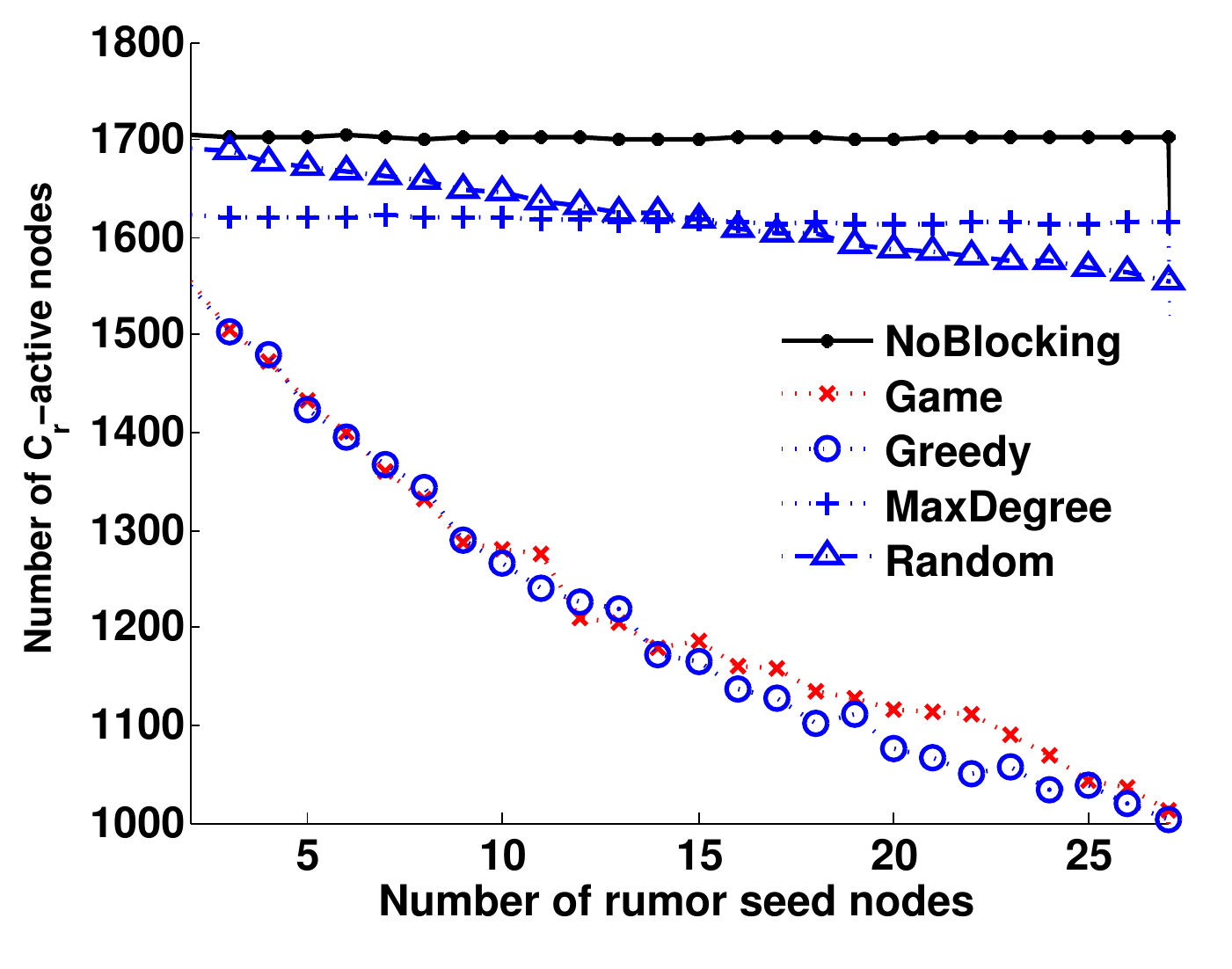}} \hspace{0mm} \ 
\subfloat[Hep with $p_{(u,v)}^G=0.1$ and $|a_r|=20$]{\label{fig:hep_01_r20}\includegraphics[width=0.32\textwidth]{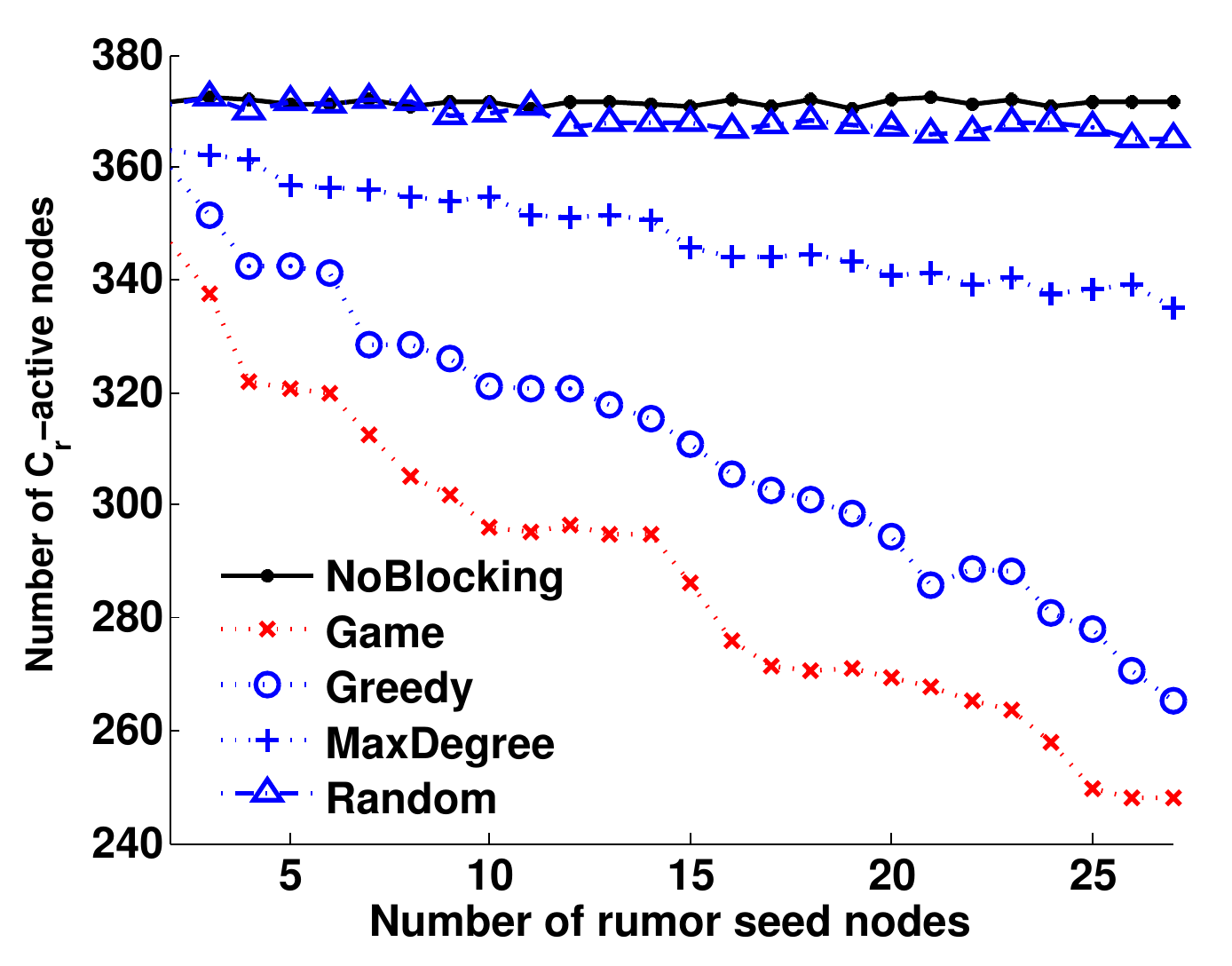}} \hspace{0mm} 
\subfloat[Facebook with $p_{(u,v)}^G=0.01$ and $|a_r|=5$]{\label{fig:facebook_001_r5}\includegraphics[width=0.32\textwidth]{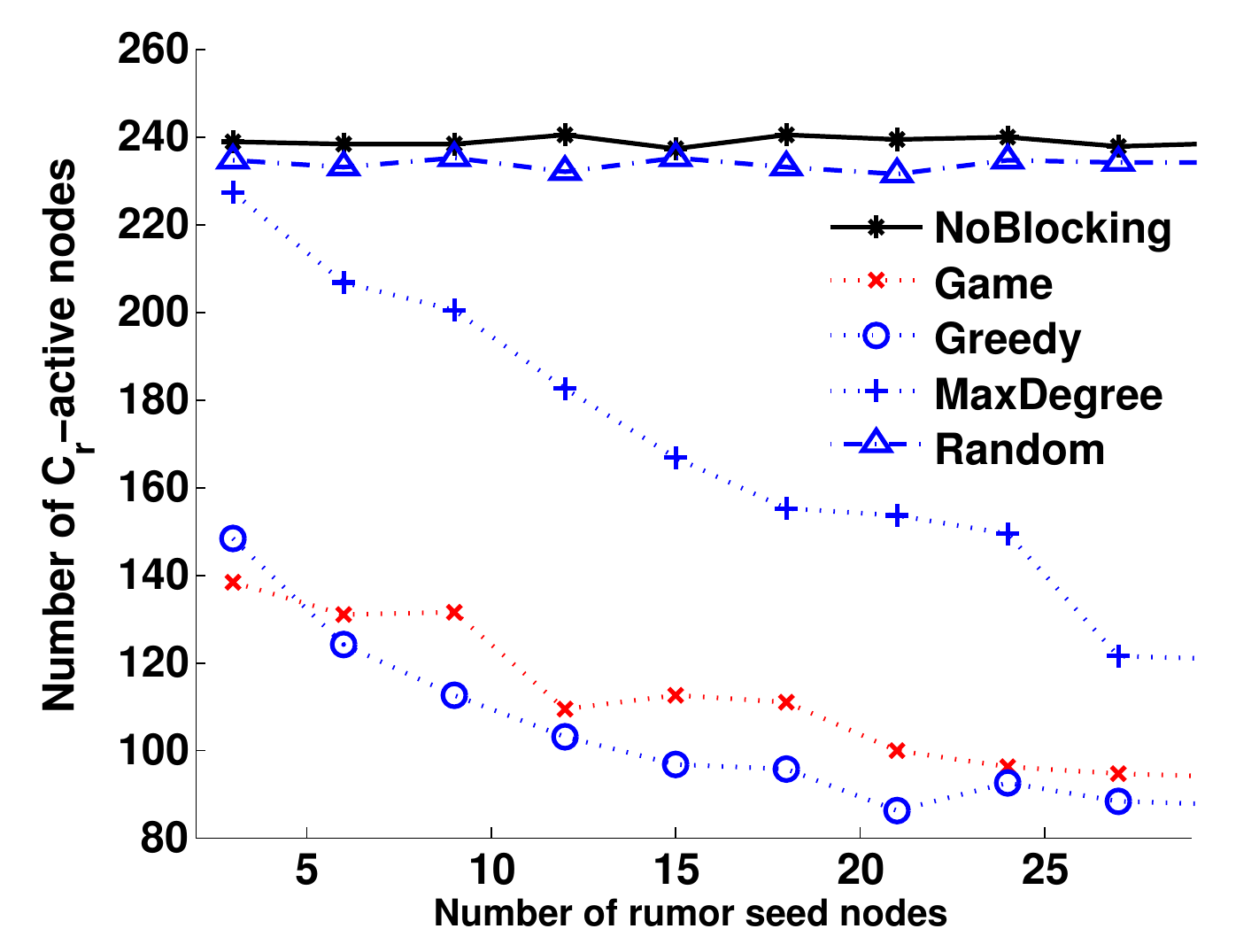}} \hspace{0mm} 

\caption{Results of the second experiment. The y-axis and x-axis denote the expected number of $C_r$-active nodes and the number of rumor seed nodes, respectively.  Each graph gives
five curves plotting the number of $C_r$-active nodes under NoBlocking, Game, Greedy, MaxDegree, and Random, respectively.} 
\label{fig:exp2}
\end{figure*}

\begin{figure*}[pt]
\centering


\subfloat[Facebook with $p_{(u,v)}^G=0.1$, $|a_r|=10$ and $k=15$ ]{\label{fig:facebook_01_round}\includegraphics[width=0.32\textwidth]{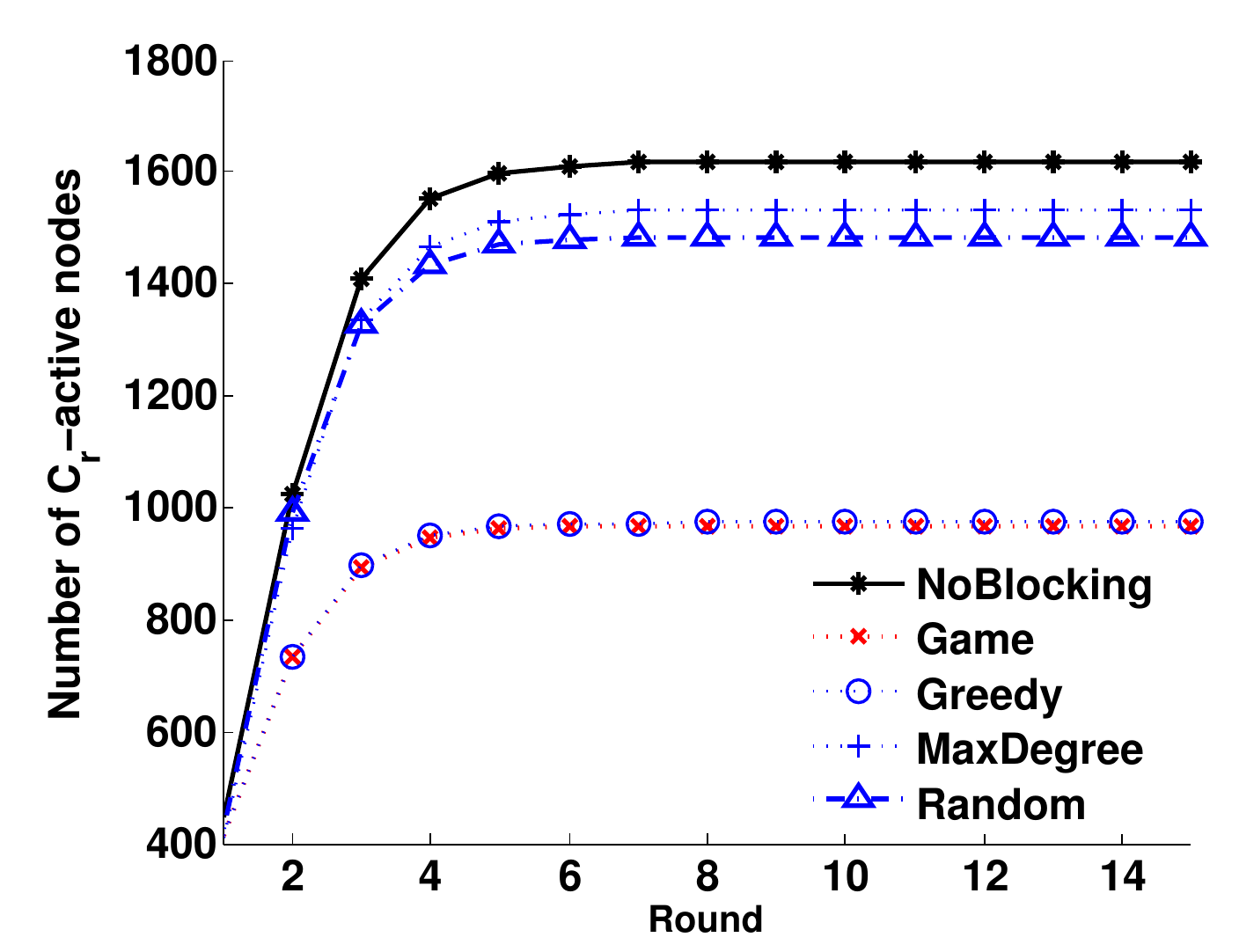}} \hspace{0mm}
\subfloat[Hep under weighted cascade setting, $|a_r|=10$ and $k=25$]{\label{fig:hep_cas_round}\includegraphics[width=0.32\textwidth]{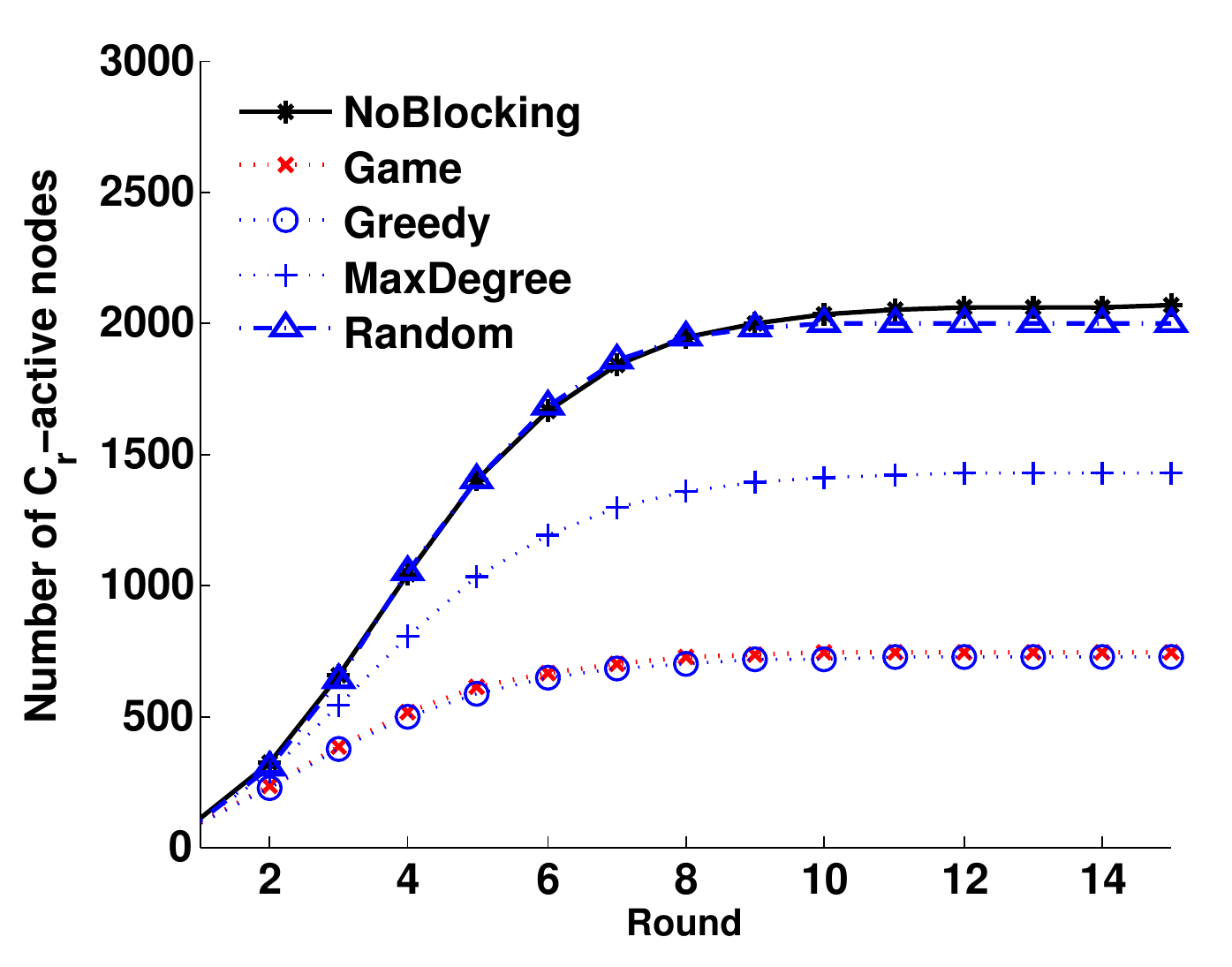}} \hspace{0mm}
\subfloat[Facebook under weighted cascade setting, $|a_r|=10$ and $k=10$]{\label{fig:facebook_cas_round}\includegraphics[width=0.32\textwidth]{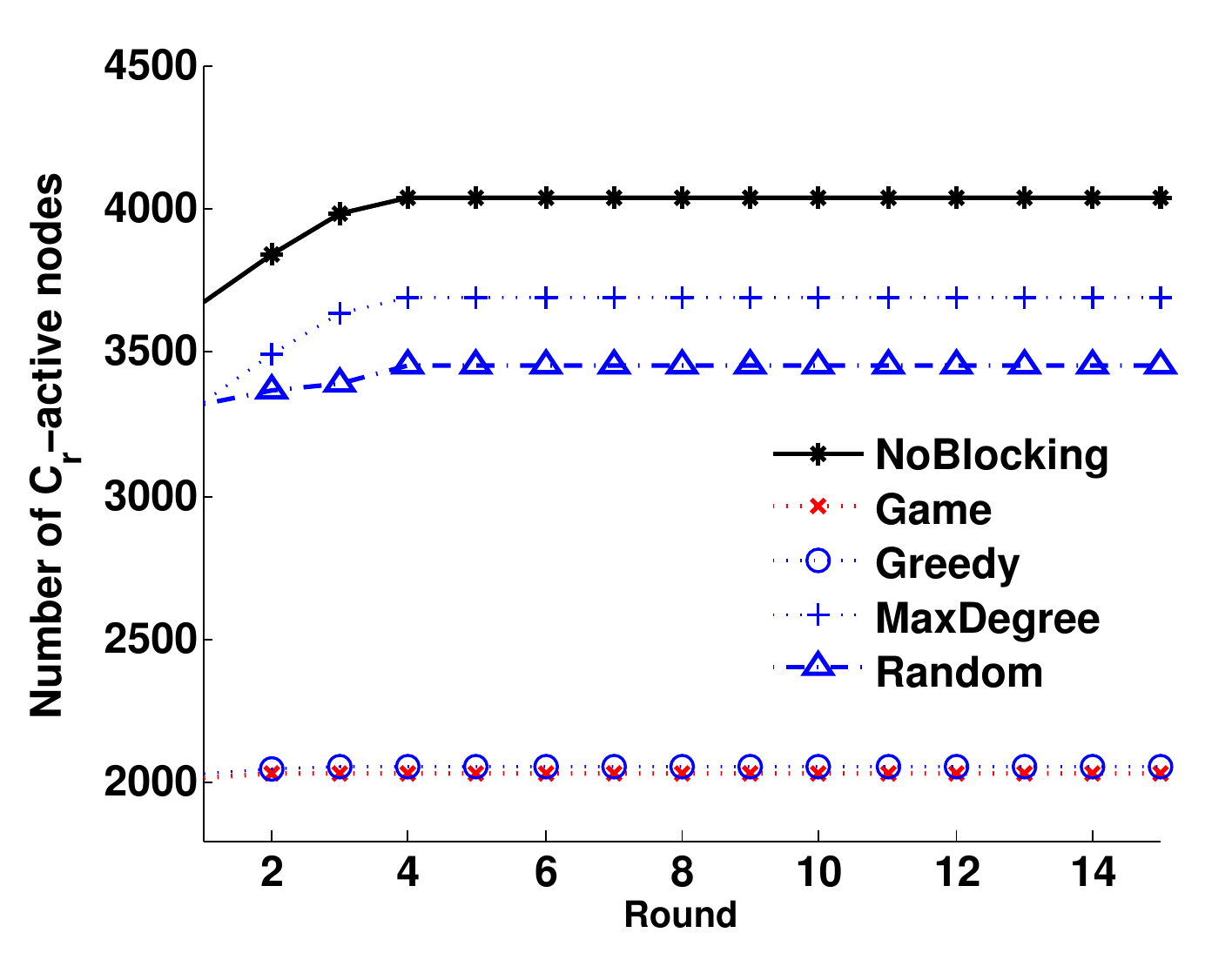}} \hspace{0mm}
\caption{Results of the third experiment. The y-axis and x-axis denote the expected number of $C_r$-active nodes and the index of spread round, respectively. Each graph gives
five curves plotting the number of $C_r$-active nodes under NoBlocking, Game, Greedy, MaxDegree and Random, respectively.} 
\label{fig:exp3}
\end{figure*}

\section{Experiment}
\label{sec:exp}
In this section, we experimentally evaluate the rumor blocking effect in the equilibrium of the proposed game by comparing it with the traditional rumor blocking framework where there is only one positive cascade. We first simulate the behavior of the agents to obtain the seed nodes, and then simulate the diffusion process to see how many users will be influenced by rumor. 

\subsection{Setup}
Our experiments are performed on a server with 16 GB ram and a 3.6 GHz quadcore processor running 64-bit JAVA VM 1.6. 

\subsubsection{Dataset} We adopt the network structure of the following datasets. The first dataset, denoted by \textbf{Facebook}, is collected from the Facebook social platform, provided by SNAP \cite{snapnets}. The Facebook dataset contains 4,039 nodes with 88,234 edges and it has been widely used in prior works \cite{mcauley2014discovering, guo2013personalized, backstrom2014romantic}. Another real-world social network is an academic collaboration from co-authorships in physics, denoted by \textbf{Hep}. This dataset is compiled from the ``High Energy Physics-Theory" section of the e-print arXiv\footnote{http://www.arXiv.org}, including about 15,000 nodes and 58,000 edges. Hep dataset has been studied in \cite{kempe2003maximizing, chen2009efficient, zhang2014minimizing} and \cite{long2011minimizing}. 

\subsubsection{Propagation Probability} In the experiments, we consider three settings of the probability on the edges. In the first and second settings, the probability of each edge is uniformly set as 0.1 and 0.01, respectively. The third setting follows the classic weighted cascade model \cite{kempe2003maximizing} where $p_{(u,v)}^G=1/d_v$ and $d_v$ is the number of out-neighbors of node $v$. 

\subsubsection{Seeds of Rumor} The seed nodes of rumor are selected from the nodes with the highest degree. The number of the rumor seeds will be discussed later.

\subsubsection{The Game} Given a social network and budget $k$, we deploy $k$ agents each of which generates one positive cascade with one seed node. The seed nodes are obtained by simulating the Simple Game developed in Sec. \ref{sec: game}. The diffusion result of the Simple Game is labeled as \textit{Game}. 

\subsubsection{Single-positive-cascade case.}For the single-positive-cascade case, we set the budget of the seed set as $k$ and select the seed nodes of the positive cascade according to the following methods. 
\begin{itemize}
\item \textit{Greedy}. This is the state-of-the-art rumor blocking algorithm. Given a budget $k$, we assume there is one positive cascade with $k$ seed nodes in which the nodes are decided by successively adding the node that can maximize the social value. Such a method provides a $(1-1/e)$-approximation due to the submodularity and it has been wildly used in the prior works \cite{budak2011limiting,fan2013least,he2012influence}. 
\item \textit{MaxDegree}. Assuming there is one positive cascade, MaxDegree selects the $k$ users in $V\setminus a_r$ with the highest degree.
\item \textit{Random}. Assuming there is one positive cascade, Random selects $k$ seed nodes at random.
\item \textit{NoBlocking}. This is the case when there is no positive cascade. 
\end{itemize}

Another popular heuristic rumor blocking algorithm, called Proximity, which selects the neighbor of rumor seed nodes as positive seed nodes, is not included in our experiments because its performance is worse than that of Greedy as shown in \cite{he2012influence} and \cite{fan2013least}. Due to space limitation, we will not discuss all combinations of the above settings. For a given full-action (i.e., the seed sets of each cascade) and a specified network, $\gamma(A)$ is calculated by taking the average of 10,000 simulations. 

\subsection{Results}
We perform three series of experiments. The experimental results are discussed as follows.

\subsubsection{Experiment \Romannum{1}} In the first experiment, the number of seed nodes of the rumor is set from 1 to 30 and the budget $k$ is equal to the number of rumor seed nodes. The results of this experiment on Facebook under the three propagation probability settings are shown in Figs. \ref{fig:facebook_001_1_30}, \ref{fig:facebook_01_1_30} and \ref{fig:facebook_cas_1_30}. As shown in the figures, when the propagation probability is 0.01, the effectiveness of Game is slightly worse than that of Greedy. Under the other two settings of propagation probability, the equilibrium of the game has the same degree of effect as Greedy does in limiting the spread of the rumor. The result of this experiment done on Hep is shown in Fig. \ref{fig:hep_01_1_30}. In this case, Game provides the best performance among all the considered methods. In general, both of the Game and Greedy are effective for rumor containment. However, under different settings and network structures, the patterns of the curves are diverse. The first observation is that the spread of rumor may have a saturation point with respect to the number of seed nodes. For the cases shown in Figs. \ref{fig:facebook_001_1_30} and \ref{fig:facebook_01_1_30}, the number of $C_r$-active nodes under NoBlocking will not notably increase when $k$ is larger than 5. Nevertheless, in Fig. \ref{fig:hep_01_1_30}, the number of  $C_r$-active nodes continuously increases with the increase of $k$. Another observation is that when $k$ increases by one, the number of $C_r$-active nodes does not necessarily increase. As shown in Fig. \ref{fig:facebook_cas_1_30}, when one rumor seed and one agent are added at $k=15$, the number of $C_r$-active nodes decreases by about 500 under Game. Such a case suggests that the marginal effect of adding one seed node not only depends on the selection of the seed nodes but also on the network structure.

\subsubsection{Experiment \Romannum{2}} In the second experiment, we fix the number of rumor seed nodes and see how the number of $C_r$-active nodes varies with the increase of the budget $k$. The results of the experiments under three different settings are shown as Figs. \ref{fig:facebook_01_r20}, \ref{fig:hep_01_r20}  and \ref{fig:facebook_001_r5}. One can see that when more budget is added the number of $C_r$-active nodes become less and less. On the Facebook network, when $p_{(u,v)}^G$ is equal to 0.1 and $|a_r|$ is set as 20, as shown in Fig. \ref{fig:facebook_01_r20}, the number of $C_r$-active nodes decreases significantly under Game and Greedy but hardly changes under MaxDegree and Random, which implies that the agent should not arbitrarily select seed nodes or use simple heuristics. For the case shown in Fig. \ref{fig:facebook_001_r5}, one can see that adding the first positive seed node can reduce the number of $C_r$-active nodes by a half. Such a scenario answers the submodularity nature of the rumor blocking problem and indicates that the first several actions of the agents are important.  

\subsubsection{Experiment \Romannum{3}} In the third experiment, we fix both the number of rumor seeds and budget $k$, and record the number of $C_r$-active nodes round by round. That is, we take the snapshots of the first two experiments and examine how fast the rumor spread under different cases. The results are shown in Figs. \ref{fig:facebook_01_round}, \ref{fig:hep_cas_round} and \ref{fig:facebook_cas_round}. As indicated by the figures, the equilibrium of the game formulated in this paper can effectively limit the spread of the rumor.  Furthermore, under appropriate strategies, the rumor can be blocked at an early stage. For example, in Fig. \ref{fig:hep_cas_round}, the number of $C_r$-active nodes stops increasing at about the eighth round under Game. However, it increases until the eleventh round under MaxDegree.

\section{Conclusion and Future Work}
\label{sec:con}
In this paper, we study the rumor blocking problem when there are multiple positive cascades. By formulating the rumor-aware game and the rumor-oblivious game, we have shown the that under the best-response and the approximate-response, the equilibrium the game provides a $2$-approximation and  $\frac{2e+1}{e+1}$-approximation, respectively, with respect to the social utility, i.e., the number of nodes that are not influenced by rumor. The theoretical results herein are well supported by the experiments done on real-world networks.

As shown in this paper, the rumor containment in a distributed mode is effective for rumor blocking. Therefore, it is interesting to design rumor blocking strategies for multiple positive cascades with the concern of the cascade priority. Another direction of the future work is to study the pure Nash equilibrium of the rumor-oblivious game. In particular, it is interesting to study the circumstance under which the pure Nash equilibrium exists. Finally, as discussed in Sec. \ref{sec:further}, the competitive cascade model becomes evasive under certain settings. To the best our knowledge, none of the prior works has considered the rumor blocking problem in such models. We leave this part for the future work.



%

\appendices

\section{Proofs}
\label{sec: proof}

\subsection{Proof of Theorem \ref{theorem:equivalent}}
\label{sec:proof_theorem_equivalent}
\begin{proof}
For each cascade $C$, we denote by $C^t$ be the set of $C$-active nodes after time step $t$. Now let us consider the spreading result after time step $t+1$. To prove the theorem, it is sufficient to show that given the $C^t$, for each cascade $C$, and $b_u^t$ for each $u$ at time step $t$, the distributions of the spread result after time step $t+1$ are the same under the two spread processes. In particular, because the cascade priorities are the same under the both process, it suffices to prove that, for any inactive node $u^{*}$ and any active node $v^{*}$, the probability that $v^{*}$ successfully actives $u^{*}$ at time step $t+1$ under the first spreading process is the same as that under the second one. 

\textbf{Process a.} An active node $v^{*}$ successfully activates the inactive node $u^{*}$ at time step $t+1$ if and only if $u^{*} \in b_{v^{*}}^t$, $v^{*}$ select $u^{*}$ to activate and the activation is successful. By Def. \ref{def:activation order}, this probability is $\frac{p_{(v^{*},u^{*})}}{|b_{v^{*}}^t|}$.

\textbf{Process b.} According to Def. \ref{def:realization}, $v^{*}$ will try to activate $u^{*}$ at time step $t+1$ if and only if  $\alpha_{v^{*}}^g(u^{*})=d_{v^{*}}-|b_{v^{*}}^t|+1$ and $p_{(v^{*},u^{*})}^g=1$. Since each permutation has the same probability to be generated, for each node $u \in b_{v^{*}}^t$, $\alpha_{v^{*}}^g(u)=d_{v^{*}}-|b_{v^{*}}^t|+1$ happens with the same probability. Therefore, for the node $u^{*}$, with the probability of $\frac{1}{|b_{v^{*}}^t|}$, $v^{*}$ will try to activate $u^{*}$ at time step $t+1$. Thus, under this process, that probability that $v^{*}$ successfully activates $u^{*}$ at time step $t+1$ is still $\frac{p_{(v^{*},u^{*})}}{|b_{v^{*}}^t|}$.
\end{proof}

\subsection{Proof of Lemma \ref{lemma:set_function}}
\label{sec:proof_lemma_set_function}

We introduce some preliminaries before proving Lemma. \ref{lemma:set_function}. For a fixed realization $g$ and a full-action $A=(a_1,...,a_k)$, the outcome of the influence diffusion in $g$ under $A$ is determined. Let $t_A^g(u)$ be the time that $u$ becomes active in $g$ under $A$.

\begin{lemma}
\label{lemma:1}
For two nodes $u_1$ and $u_2$, and any simple path $P=(v_1=u_1,...,v_i,...,v_m=u_2)$ from $u_1$ to $u_2$ in $g$, $t_A^g(u_2) \leq t_A^g(u_1)+|P|$ where $|P|$ is the sum of the weights of the edges in $P$. 
\end{lemma}
\begin{proof}
For any two successive node $v_{i}$ and $v_{i+1}$ in the path, $v_i$ will attempt to activate $v_{i+1}$ at $t_A^g(v_i)+w^g(v_i,v_{i+1})$. Therefore, $t_A^g(v_{i+1}) \leq t_A^g(v_i)+w^g(v_i,v_{i+1})$. The property follows inductively from $u_1$ to $u_2$ along the given path. 
\end{proof}

\begin{corollary}
\label{corollary:lemma:1}
Given a realization $g$ and a full-action $A$, for any node $u$ and $v$ where $v$ is a seed selected by one or more cascades, $t_A^g(u) \leq \mathrm{dis}^g(v,u)$.
\end{corollary}
\begin{proof}
Since $v$ is a seed node, $t_A^g(u)=0$. The corollary directly follows from Lemma \ref{lemma:1}.
\end{proof}

The next lemma provides the condition for a node $u$ to be rumor-activated in a realization $g$ under a full-action $A$. 
\begin{lemma}
\label{lemma:key}
Given a full-action $A=(a_1,...,a_k)$ and a realization $g$, a node $u^{*}$ will be activated by rumor $C_r$ in $g$ under process $b$ defined in Theorem \ref{theorem:equivalent} $\Leftrightarrow$
$\mathrm{dis}^g(a_r,u^{*}) \leq \mathrm{dis}^g(a_i,u^{*})$ for each $i$ and $\mathrm{dis}^g(a_r,u^{*}) \neq +\infty$\footnote{We define that $\mathrm{dis}^g(u_1,u_2) = \mathrm{dis}^g(u_3,u_4)$ if $\mathrm{dis}^g(u_1,u_2)=+\infty$ and $\mathrm{dis}^g(u_3,u_4)=+\infty$, for any four nodes $u_1$, $u_2$, $u_3$ and $u_4$.}.
\end{lemma}
\begin{proof}
Let $v_i$ be the node in $a_i$ such that $\mathrm{dis}^g(v_i,u^{*})=\mathrm{dis}^g(a_i,u^{*})$, for $1 \leq i \leq k$.

$\Rightarrow$: Clearly, $\mathrm{dis}^g(a_r,u^{*}) \neq +\infty$. Since $u^{*}$ is activated by rumor, there is path $P$ from a certain node $v_r \in a_r$ to $u^{*}$ such that all the nodes in this path are activated by rumor and $t_A^g(u^{*})=|P|$. By definition, $\mathrm{dis}^g(a_r,u^{*}) \leq |P|$. By Corollary \ref{corollary:lemma:1}, $t_A^g(u^{*}) \leq \mathrm{dis}^g(v_i,u^{*})$ for each $i$, which implies $\mathrm{dis}^g(a_r,u^{*}) \leq |P|= t_A^g(u^{*}) \leq  \mathrm{dis}^g(v_i,u^{*})$. 

$\Leftarrow$: Now suppose $\mathrm{dis}^g(a_r,u^{*}) \leq \mathrm{dis}^g(a_i,u^{*})$ for each $i$ and $\mathrm{dis}^g(a_r,u^{*}) \neq +\infty$. Let $v_r$ be the node in $a_r$ such that $\mathrm{dis}^g(v_r,u^{*})=\mathrm{dis}^g(a_r,u^{*})$ and $P$ be the shortest path from $v_r$ to $u^{*}$ in $g$. Suppose the nodes in $P$ are $u_1,..., u_l$ where $u_1=v_r$ and $u_l=u^{*}$. \textit{It suffices to show that every node $u_j$ in $P$ will be activated by rumor at time $\mathrm{dis}^g(u_1,u_j)$}. We prove this inductively. Clearly, $t_A^g(u_1)=\mathrm{dis}^g(u_1,u_1)=0$ and $u_1$ is activated by rumor. Suppose this is true for the first $j$ nodes in $P$ and $u_{j+1}$ is activated by its certain in-neighbor $v^{*}$. There are two cases shown as follows. 

\textbf{Case 1: $v^{*}=u_j$.} By the inductive hypothesis $u_{j}$ is activated by rumor at time $\mathrm{dis}^g(u_1,u_j)$. Therefore, $v_{j+1}$ is also activated by rumor and 
\begin{eqnarray*}
t_A^g(u_{j+1})&=&t_A^g(u_j)+w^g(u_j,u_{j+1})\\
&=&\mathrm{dis}^g(u_1,u_j)+w^g(u_j,u_{j+1})=\mathrm{dis}^g(u_1,u_{j+1}).
\end{eqnarray*}
\textbf{Case 2: $v^{*} \neq u_j$.} Suppose $v^{*}$ is activated at $t_A^g(v^{*})$ via a certain path $P^{'}$ from a certain seed node $v^{'}$ to $v^{*}$. Then, $u_{j+1}$ is activated at $|P|^{'}+w^g(v^{*},u_{j+1})$. By Corollary \ref{corollary:lemma:1},
\begin{equation}
\label{eq:pro_2_1}
|P|^{'}+w^g(v^{*},u_{j+1})=t_A^g(u_{j+1}) \leq \mathrm{dis}^g(u_1,u_{j+1}).
\end{equation}
Furthermore, since $\mathrm{dis}^g(u_1,u^{*}) = \mathrm{dis}^g(a_r,u^{*}) \leq \mathrm{dis}^g(a_i,u^{*})$, $\mathrm{dis}^g(u_1,u_{j+1}) \leq \mathrm{dis}^g(v_i,u_{j+1})$ for each $i$. Since $P^{'}$ together with $(v^{*},u_{j+1})$ is a path from $v^{'}$ to $u_{j+1}$,  $\mathrm{dis}^g(u_1,u_{j+1}) \leq |P^{'}|+w^g(v^{*},u_{j+1})$. Combining Eq.(\ref{eq:pro_2_1}),
{\small 
\begin{eqnarray*}
\mathrm{dis}^g(u_1,u_{j+1})= |P^{'}|+w^g(v^{*},v_{i+1})=t_A^g(v_{j+1}).
\end{eqnarray*}
}
By the inductive hypothesis, $u_j$ is $C_r$-active at $\mathrm{dis}^g(v_1,u_j)$ and it will attempt to activate $u_{j+1}$ at $\mathrm{dis}^g(v_1,u_j)+w^g(u_{j},u_{j+1})=\mathrm{dis}^g(u_1,u_{j+1})$, which means $u_j$ and $v^{*}$ activate $u_{j+1}$ at the same time. Since the rumor has the highest priority, $u_{j+1}$ will be activated by rumor at time step  $\mathrm{dis}^g(u_1,u_{j+1})$. By the above induction, all the nodes in path $P$, including $u^{*}$, are $C_r$-active.
\end{proof}

One can see that the minimum of $\mathrm{dis}^g(a_i,u^{*})$ only depends on the union of the positive seed sets and therefore the social utility is a set function, shown as follows. 

\begin{proof}[\textbf{Proof of Lemma \ref{lemma:set_function}}]
It suffices to show that, for any realization $g$ and two full actions $A_1$ and $A_2$, $\gamma^g(A_1)=\gamma^g(A_2)$ if $A_1^{*}=A_2^{*}$. Let $u$ be an arbitrary $\overline{C}_r$-active node in $g$ under $A_1=(a_1,...,a_k)$. By Lemma \ref{lemma:key}, $\mathrm{dis}^g(a_r,u) > \mathrm{dis}^g(a_i,u)$ for some $i$ or $\mathrm{dis}^g(a_r,u)=\infty$. If $\mathrm{dis}^g(a_r,u)=\infty$, then there is no path from any rumor seed to $u$ and therefore $u$ cannot be activated by rumor in $g$ under $A_2$. Now suppose $\mathrm{dis}^g(a_r,u) \neq \infty$ and  $\mathrm{dis}^g(a_r,u) > \mathrm{dis}^g(v_i,u)$ for some $i$ and some $v_i \in a_i$. Since $A_1^{*}=A_2^{*}$, $v_i$ must be a seed node of some agent $i^{*}$ in $A_2$ and therefore $\mathrm{dis}^g(a_r,u) < \mathrm{dis}^g(a_{i^*},u)$, which means $u$ is also $\overline{C}_r$-active in $g$ under $A_2$. By the above analysis, $\gamma^g(A_1) \leq \gamma^g(A_2)$. It can be easily seen that $\gamma^g(A_2) \leq \gamma^g(A_1)$ can be proved in the similar manner, and therefore $\gamma^g(A_1) = \gamma^g(A_2)$.
\end{proof}

\subsection{Proof of Lemma \ref{lemma:submodular}}
\label{sec:proof_lemma_submodular}
By Theorem \ref{theorem:equivalent}, we only need to show that $\gamma^g()$ is monotone increasing and submodular for each realization $g$. Due to Lemma \ref{lemma:key}, $\gamma^g()$ is clearly monotone increasing. To prove the submodularity, it suffices to show that, for each realization $g$,  
\begin{equation}
\label{eq:submodular}
\gamma^g(X \cup \{v\})-\gamma^g(X) \geq \gamma^g(Y \cup \{v\})-\gamma^g(Y),
\end{equation}
where $X \subseteq Y \subseteq V$ and $v \in V \setminus Y$. Since $\gamma^g()$ is  monotone increasing, $\gamma^g(Y \cup \{v\})-\gamma^g(Y)$ is the number of nodes that are $C_r$-active under $Y$ but $\overline{C}_r$-active under $Y \cup \{v\}$. Now let us consider such a node $u$. Since $u$ is $C_r$-active under $Y \cup \{v\}$, $\mathrm{dis}^g(a_r,u) \neq +\infty$. By Lemma \ref{lemma:key}, $\mathrm{dis}^g(a_r, u) \leq \mathrm{dis}^g(Y,u)$ and $\mathrm{dis}^g(a_r, u) > \mathrm{dis}^g(Y \cup \{v\},u)$, which means $\mathrm{dis}^g(a_r, u) > \mathrm{dis}^g(\{v\},u)$. Because $X$ is a subset of $Y$, $\mathrm{dis}^g(X,u) \geq \mathrm{dis}^g(Y,u) \geq  \mathrm{dis}^g(a_r,u)$ and thus $u$ will be $C_r$-active in $g$ under $X$. Meanwhile, $u$ cannot be $C_r$-active in $g$ under $X \cup \{v\}$ because $\mathrm{dis}^g(X \cup \{v\},u) \leq \mathrm{dis}^g(\{v\},u) < \mathrm{dis}^g(a_r,u)$. Therefore, each node that contributes 1 to the right-hand side of Eq. (\ref{eq:submodular}) must contribute 1 to the left-hand side. Eq. (\ref{eq:submodular}) thus follows.

\subsection{Proof of Lemma \ref{lemma:pre_second}}
\label{sec:proof_lemma_pre_second}
\begin{proof}
By the definition of $\overline{\delta}_i(\cdot)$, Eq. (\ref{eq:utility}) directly follows. Let $\delta_i^g(A)=\gamma^g(A)-\gamma^g(A(i,\emptyset))$. To prove Eq. (\ref{eq:valid}), it suffices to show that $\sum_{i=1}^{k} \delta_i^g(A) \leq \gamma^g(A)$ holds for each full-action $A$ and $g$. Note that
{\small 
\begin{eqnarray*}
&&\sum_{i=1}^{k}\delta_i^g(A)\\
&=& \sum_{i=1}^{k}\gamma^g(A)-\gamma^g(A(i,\emptyset)) \\
&=& \sum_{i=1}^{k}\gamma^g(A^k)-\gamma^g(A^k(i,\emptyset)) \\
&&\{\text{By~Corollary~\ref{theorem: submodular}}\}\\
&\leq& \sum_{i=1}^{k}\gamma^g(A^{i})-\gamma^g(A^{i}(i,\emptyset)) \\
&=&\gamma^g(A^k)=\gamma^g(A).
\end{eqnarray*} 
}
Thus, proved.
\end{proof}

\subsection{Proof of Theorem \ref{theorem:rumor_aware_valid}}
\label{sec:proof_theorem_rumor_aware_valid}
\begin{proof}
Suppose $\Omega=(b_1,...,b_k)$. For two node sets $V_1$ and $V_2$, we denote by $V_1^{V_2}$ the set of the nodes in $V_1$ but not in $V_2$, i.e., $V_1^{V_2}=V_1-V_2$. Under such notation, $A^{*} \cup \Omega^{*}=A^{*} \cup b_1^{A^{*}}\cup,...,\cup b_k^{A^{*}}$. Let $B_i=\{b_1^{A^{*}},...,b_i^{A^{*}}\}$ and $B_0=\{\emptyset\}$. Then 
\begin{eqnarray*}
&&\gamma(A^{*} \cup \Omega^{*})-\gamma(A^{*}) \\
&=&\sum_{i=1}^{k} \gamma(A^{*} \cup B_i^*)-\gamma(A^{*} \cup B_{i-1}^*). 
\end{eqnarray*}
Due to submodularity, for each $1 \leq i \leq k$, 
\begin{eqnarray*}
&& \gamma(A^{*} \cup B_i^*)-\gamma(A^{*} \cup B_{i-1}^*) \\
&\leq& \gamma(A^{*} \cup b_i^{A^{*}})-\gamma(A^{*}) \\
&\leq& \gamma(A^{*}-a_i \cup b_i^{A^{*}})-\gamma(A^{*}-a_i) \\
\end{eqnarray*}
According to the monotonicity of $\gamma()$, 
\begin{eqnarray*}
\gamma(A^{*}-a_i \cup b_i^{A^{*}}) \leq \gamma(A^{*}-a_i \cup b_i).
\end{eqnarray*}
Therefore,
\begin{eqnarray}
&&\gamma(A^{*} \cup \Omega^{*})-\gamma(A^{*})  \nonumber \\ 
&\leq&\sum_{i=1}^{k} \gamma(A^{*}-a_i \cup b_i^{A^{*}})-\gamma(A^{*}-a_i)  \label{eq:1} \nonumber\\  
&\leq&\sum_{i=1}^{k} \gamma(A^{*}-a_i \cup b_i)-\gamma(A^{*}-a_i) \\
&&\{\text{by the approximate response}\}  \nonumber \\  
&\leq&\sum_{i=1}^{k} (1-e^{-1})^{-1} \cdot \delta_i(A^{*}) \nonumber \\ 
&&\{\text{by Eq. (\ref{eq:valid})}\} \nonumber \\  
&\leq&(1-e^{-1})^{-1} \cdot \gamma(A^{*}). \label{eq:2}
\end{eqnarray}
Thus, proved. 
\end{proof}

\subsection{Proof of Lemma \ref{lemma:rumor_oblivious_valid_utility}}
\label{sec:proof_lemma:rumor_oblivious_valid_utility}
Let $\sigma_i^g(A)$ be the number of $C_i$-active nodes in $g$, By Theorem \ref{theorem:equivalent},
\begin{equation*}
\sigma_i(A)=\sum_{g \in \mathcal{G}} \mathrm{Pr}[g] \cdot \sigma_i^g(A),
\end{equation*}
and therefore, in order to prove Eq. (\ref{eq:utility}), it suffices to show that $\sigma_i^g(A) \geq \gamma^g(A)-\gamma^g(A(i,\emptyset))$ holds for each full-action $A$ and each realization $g$. Note that $\gamma^g(A)$ is a set function  on $A^{*}$ and it is monotonically increasing. Therefore, the right-hand side is number of nodes that are $C_r$-active under $A$ but $\overline{C}_r$-active under $A(i,\emptyset)$. Let $u$ be such a node that contribute 1 to the right-hand side. According to Lemma \ref{lemma:key}, $\mathrm{dis}^g(a_i,u) < \mathrm{dis}^g(a_r,u) \leq \mathrm{dis}^g(a_j,u)$ for $j \neq i$. Thus, $u$ must be $C_i$-active under $A$ in $g$ and therefore also contributes 1 to the left-hand side. Eq. (\ref{eq:utility}) thus proved.

Now to prove Eq. (\ref{eq:valid}), it suffices to show that $\sum_{i=1}^{k} \delta_i^g(A) \leq \gamma^g(A)$ holds for each $A$ and $g$. Note that $\gamma^g(A)$ is the number of $\overline{C}_r$-active nodes, i.e., the nodes activated by the positive cascades together with the nodes that are not activated by any cascade. Therefore, Eq. (\ref{eq:valid}) follows directly.




\ifCLASSOPTIONcaptionsoff
  \newpage
\fi



\bibliographystyle{IEEEtran}
\bibliography{sigproc}
%



%

\begin{IEEEbiography}[{\includegraphics[width=1in,clip,keepaspectratio]{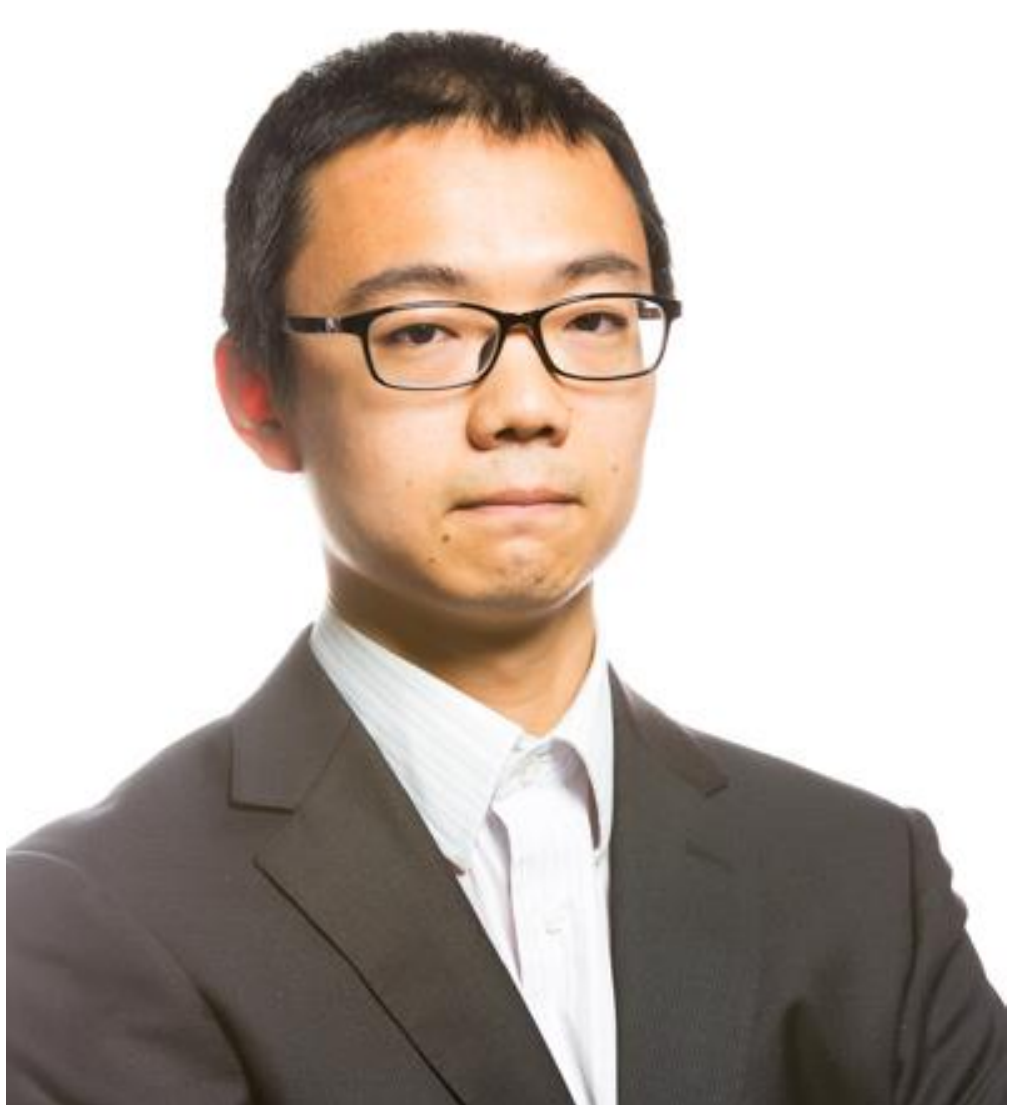}}]{Amo (Guang-mo) Tong} is a Ph.D candidate in the Department of Computer Science at the University of Texas at Dallas under the supervision of Dr. Ding-Zhu Du. He received his BS degree in Mathematics and Applied Mathematics from Beijing Institute of Technology in July 2013. His research interests include computational social system, bigdata analysis and real-time systems. He has published several papers on prestigious conferences and journals.  He is a student member of the IEEE.
\end{IEEEbiography}
\begin{IEEEbiography}[{\includegraphics[width=1in,clip,keepaspectratio]{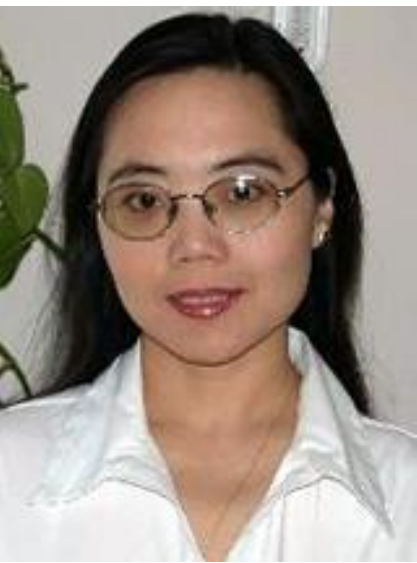}}]{Weili Wu}(M'00) is currently a Full Professor with the Department of Computer Science, University of Texas at Dallas, Dallas, TX, USA. She received the Ph.D. and M.S. degrees from the Department of Computer Science, University of Minnesota, Minneapolis, MN, USA, in 2002 and
1998, respectively. Her research mainly deals in the general research area of data communication and data management. Her research focuses on the design and analysis of algorithms for optimization problems that occur in wireless networking environments and various database systems.
\end{IEEEbiography}
\begin{IEEEbiography}[{\includegraphics[width=1in,clip,keepaspectratio]{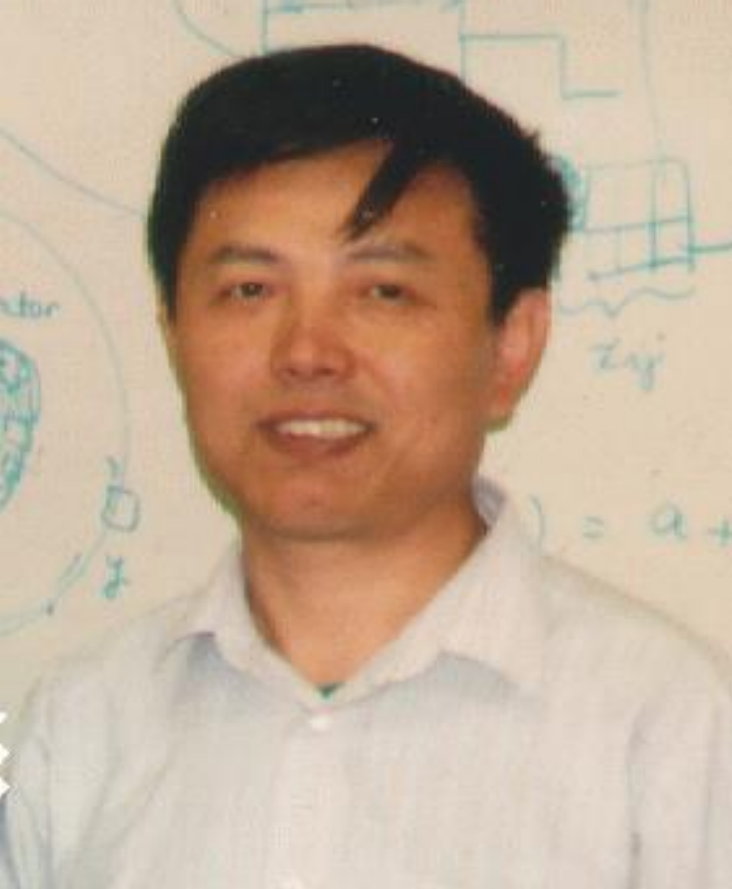}}]{Ding-Zhu Du} received the M.S. degree from the Chinese Academy of Sciences in 1982 and the Ph.D. degree from the University of California at Santa Barbara in 1985, under the supervision of Professor Ronald V. Book. Before settling at the University of Texas at Dallas, he worked as a professor in the Department of Computer Science and Engineering, University of Minnesota. He is the editor-in-chief of the Journal of Combinatorial Optimization and is also on the editorial boards for several other journals. Forty Ph.D. students have graduated under his supervision. He is a member of the IEEE
\end{IEEEbiography}




\vfill


\end{document}